\documentclass[12pt,article]{iopart}
\usepackage{epsfig}
\renewcommand{\vec}[1]{\mbox{\boldmath $ #1 $}}
\newcommand{\grad}{{\mbox{\boldmath$\nabla$}}}
\newcommand{\rot}{\mbox{\boldmath $\nabla \times $}}
\newcommand{\eqref}[1]{(\ref{#1})}
\newcommand{\be}{\begin{equation}}
\newcommand{\ee}{\end{equation}}
\newcommand{\beq}{\begin{equation}}
\newcommand{\eeq}{\end{equation}}
\newcommand{\ber}{\begin{eqnarray}}
\newcommand{\eer}{\end{eqnarray}}
\newcommand{\bea}{\begin{eqnarray}}
\newcommand{\eea}{\end{eqnarray}}
\renewcommand{\lsim}{ {{}_{<}\atop{{}^{\sim}}} }
\newcommand{\gsim}{ {{}_{>}\atop{{}^{\sim}}} }

\newcommand{\vDelta}{\mbox{\boldmath$\slDelta$}}
\newcommand{\slPhi}{{\it \Phi} }
\newcommand{\slDelta}{{\it \Delta} }
\begin{document}
%\title{
%\vspace*{-1cm}{\small\bf Invited Paper for the Focus Volume on "Superconductors with Exotic Symmetries" for New Journal of Physics}\vspace*{0.8cm}\\
\title{Charge Dynamics of Vortex Cores in Layered Chiral Triplet Superconductors}
\author{M. Eschrig}
\address{Institut f{\"u}r Theoretische Festk{\"o}rperphysik
and DFG-Center for Functional Nanostructures,
Universit{\"a}t Karlsruhe, D-76128 Karlsruhe, Germany}
\ead{\mailto{eschrig@tfp.uni-karlsruhe.de}}
\author{J.A. Sauls}
\address{Department of Physics \& Astronomy, Northwestern University, Evanston, IL 60208, USA}
\ead{\mailto{sauls@northwestern.edu}}
\date{\today}
%-----------------------------------------------------------------------------------------------------------
\begin{abstract}
In an accompanying paper 
[J.A. Sauls and M. Eschrig, {\it Vortices in chiral, spin-triplet superconductors and superfluids}, 
arXiv:0903.0011] we have studied the equilibrium properties of vortices in a chiral quasi-twodimensional triplet super\-fluid/super\-conductor.
Here we extend our studies to include the dynamical response of a vortex core in a chiral triplet superconductor to an external {\em a.c.} electromagnetic field. 
We consider in particular the response of a doubly quantized vortex with a 
homogeneous core in the time-reversed phase.
The external frequencies are assumed to be comparable in magnitude to the
superconducting gap frequency, such that the vortex motion is 
non-stationary but can be treated by linear response theory.
We include broadening of the vortex core bound states due to impurity scattering and
consider the intermediate clean regime, with a broadening comparable to or 
larger than the quantized energy level spacing. 
The response of the order parameter, impurity self energy,
induced fields and currents are obtained by a self-consistent calculation
of the distribution functions and the excitation spectrum.
Using these results we obtain the self-consistent
dynamically induced charge distribution in the vicinity
of the core. This charge density is related to the nonequilibrium
response of the bound states and order parameter collective mode, and
dominates the electromagnetic response of the vortex core.
\end{abstract} 
%-----------------------------------------------------------------------------------------------------------
\maketitle
\tableofcontents
\markboth{Charge Dynamics of Vortex Cores in Layered Chiral Triplet Superconductors}{Charge Dynamics of Vortex Cores in Layered Chiral Triplet Superconductors}
%-----------------------------------------------------------------------------------------------------------
\section{Introduction}

Transport and optical properties in type II superconductors are
closely related to the response of Abrikosov vortices \cite{aaa57} 
to external electromagnetic fields. 
Abrikosov vortices that form a lattice in sufficiently strong magnetic fields
give rise to a finite resistance when a voltage is applied to the
superconducting sample.  This flux flow resistance appears,
once collective vortex pinning \cite{lar79} is overcome, 
as a result of motion of the vortex lattice in an electric field 
\cite{degen64} with a resistance independent on the field strength for
small fields \cite{gor73,LO73,GK75}.
In superconductors with moderate impurity disorder the vortex lattice
moves in a constant electric field with constant velocity $\vec{v}_{\rm L}$
predominantly in the direction transverse to the transport current, 
\begin{equation}
\vec{v}_{\rm L} = \frac{\mu }{c} \vec{j}_{\rm tr} \times \vec{\Phi}_0
\end{equation}
with 
$\vec{\Phi}_0 = (hc/2|e|) \, \vec{n}_B $ being the magnetic flux quantum
multiplied with the unit vector
$\vec{n}_B=\vec{B}_{\rm av}/B_{\rm av}$ that describes the direction of the 
averaged over the vortex unit cell magnetic field , $\vec{B}_{\rm av}$, and
$\mu $ quantifies the vortex mobility.
This viscous flux flow is in contrast to superclean superconductors
or to neutral superfluids, where vortices move predominantly 
together with the superfluid velocity.
The electric field averaged over the unit cell, $\vec{E}_{\rm av} $
is given by
\begin{equation}
\vec{E}_{\rm av} = \frac{1}{c} \vec{B}_{\rm av} \times \vec{v}_{\rm L} .
\end{equation}
On a phenomenological level, vortex motion has been discussed within the
Bardeen-Stephen model \cite{bar65}, the Nozi{\`e}res-Vinen model
\cite{noz66}, and within time-dependent Ginzburg-Landau theory
\cite{sch66}.
The microscopic theory for flux flow resistance has been pioneered 
by  Gor'kov and Kopnin \cite{gor73,GK75} and by
Larkin and Ovchinnikov \cite{lar79,LO73,LO75b,LO76,lar86}.
In extremely clean superconductors vortices move predominantly
in direction of the transport
current, giving rise to an electric field perpendicular to the transport current,
and thus a Hall effect \cite{kop76,dor92,kop93,kop94,kop95,LO95,kop96,hou98}. 

Vortices also provide valuable information about the
nature of low lying excitations in the superconducting state.
In clean $s$-wave BCS superconductors the low-lying
excitations in the core are the bound states of
Caroli, de Gennes and Matricon \cite{car64}.
These excitations have superconducting as well as normal
metallic properties. For example, these states are the
source of circulating supercurrents in the equilibrium vortex core,
and they are strongly coupled to the condensate by Andreev scattering
\cite{bar69,rai96}. Furthermore, the response of the vortex core states
to an electromagnetic field is generally very different
from that of normal electrons.

In the following we will discuss the response of a vortex to an
{\em a.c.} electromagnetic field with frequency $\omega $. 
We are interested in frequencies $\omega $
that are comparable to the frequencies set by the
superconducting gap, $\slDelta /\hbar $, i.e. $1\,\mbox{GHz}-1\,\mbox{THz}$. In this region
the vortex motion is non-adiabatic and is not described by the
low-frequency limit of flux flow motion. 
We consider {\sl layered} superconductors with a weak Josephson coupling between the 
conducting planes. In this case the vortices are quasi-two-dimensional objects, called
{\sl pancake vortices}. The peculiarities of the vortex phases in layered superconductors
has been reviewed in Ref. \cite{Blatter94} (for recent
developments see also \cite{Bill_vortex}).

Disorder plays a central role in the dissipative dynamics of the
mixed state of type II superconductors. Impurities and defects are a source
of scattering that limits the mean free path, $\ell$, of carriers, thus increasing
the resistivity. Defects also provide `pinning sites' that inhibit
vortex motion and suppresses the flux-flow resistivity \cite{lar79}.
However, for {\em a.c.} fields even pinned vortices are sources for dissipation.
The magnitude and frequency dependence of this dissipation depends on the 
electronic structure and dynamics of the core states of the pinned vortex.
Experimental studies of dissipation in vortex cores include Refs.
\cite{karrai92,eldridge95,mae07}.

When discussing vortex cores states in the context of impurity disorder,
one needs to distinguish three degrees of disorder, which we summarize
in Table \ref{t1}. 
%
%\begin{center}
\begin{table}[b]
\caption{\label{t1}
Three degrees of disorder are distinguished for vortices in superconductors with impurities. Here, $1/\tau =v_f/\ell $, with the normal state 
Fermi velocity $v_f$.}
\begin{indented}
\item[]\begin{tabular}{|c|c|c|}
\hline
superclean case & moderately clean & dirty limit \\
\hline 
\hline &&\\[-0.3cm]
$\displaystyle \ell \gg \xi_0 \frac{E_f}{\slDelta } $ &  
$\displaystyle \xi_0\frac{E_f}{\slDelta } \gg \ell \gg \xi_0 $ &  
$\displaystyle \xi_0 \gg \ell \gg \xi_0 \frac{\slDelta }{E_f } $ 
\\[0.3cm]
\hline &&\\[-0.3cm]
$\displaystyle \frac{\hbar}{\tau }\ll \frac{\slDelta^2 }{E_f} $ &  
$\displaystyle \frac{\slDelta^2 }{E_f} \ll \frac{\hbar}{\tau } \ll \slDelta  $ &  
$\displaystyle \slDelta   \ll \frac{\hbar}{\tau }  \ll  E_f  $ 
\\[0.3cm]
\hline
\end{tabular}
\end{indented}
\end{table}
%\end{center}
%
In the dirty limit, $\hbar/\tau\gg\slDelta$, the the Bardeen-Stephen
model \cite{bar65} of a normal-metal spectrum with the
local Drude conductivity in the core
provides a reasonable description of the dissipative dynamics of the vortex core.
The opposite extreme is the ``superclean limit'',  $\hbar/\tau\ll\slDelta^2/E_f$
(where $E_f$ is the Fermi energy),
in which the quantization of the vortex-core bound states must be taken into account. 
In this limit a single impurity and its interaction with the vortex core states
must be considered. The {\em a.c.} electromagnetic response is then controlled
by selection rules governing transition matrix elements for the quantized
core levels and the level structure of the core states in the presence
of an impurity \cite{jan92,zhu93,hsu95,lar98,kou99,atk99}.
In the case of d-wave superconductors in the superclean limit, 
{\sl minigap nodes}
in the spectrum of bound states lead to a finite dissipation from Landau
damping for $T\rightarrow 0$ \cite{kop97}.

The superclean limit is difficult to achieve even for short coherence
length superconductors; weak disorder broadens the vortex
core levels into a quasi-continuum. We investigate the intermediate-clean
regime, $\slDelta^2/E_f\ll\hbar/\tau\ll\slDelta$, where the discrete
level structure of the vortex-core states is broadened and the selection
rules are broken {due to strong overlap between the bound state wave functions}.
However, the vortex core states remain well defined on the
scale of the superconducting gap, $\slDelta $. In this regime we can
take advantage
of the power of the quasiclassical theory of nonequilibrium
superconductivity \cite{eil68,lar68,eli71,schmid75,LO75b}.
Previous results for $s$-wave and $d$-wave vortices in
layered superconductors \cite{esc97,esc99,esc01,esc02} have shown
that in the the intermediate-clean regime electrodynamics
of the vortex state is nonlocal and
largely determined by the response of the vortex-core states.
Transitions involving the vortex-core states, and their coupling to the
collective motion of the condensate requires dynamically self-consistent
calculations of the order parameter, self energies, induced fields,
excitation spectra and distribution functions.
In particular, it has been found that branches of localized
vortex bound states that cross the chemical potential  (so-called
anomalous branches)
are of crucial importance for the dissipation
in the vortex core \cite{esc99}. The relaxation of localized excitations
takes place via their interaction with impurities.
The importance of anomalous branches for the flux flow resistance 
in moderately clean superconductors
has been clarified by Kopnin and Lopatin \cite{kop95}.

An interesting aspect of vortex dynamics is the charge dynamics associated
with moving vortices. It is known that even for a static vortex
charge transfer takes place from the vortex core region to
outer regions of the vortex, leading
to a depletion of charge carriers in the vortex core regions
\cite{LO95,kho95,fei95,bla96,kol01,kum01,che02,shi02,lip02,mac03,kna05,zhu06,zha08}.
Associated with this charge transfer is a spatially varying 
electrostatic potential that is typically of the order of $\slDelta^2/E_f$. 
The smallness of the parameter $\slDelta/E_f$ means that the electrostatic 
potential can be neglected in the self-consistent determination of the spatially 
varying order parameter and magnetic field.

There are two main sources for vortex charges.
The first contribution comes from the lowering of free energy by the pairing, 
which leads to a force on the unpaired electrons in the vortex core by the
condensate around the vortices \cite{Sorokin49}. 
This force varies on the coherence length scale.
It arises due to the change of the chemical potential as function of
the modulus of the order parameter, and is in general proportional to 
$\partial \gamma /\partial n$ and $\partial T_{\rm c}/\partial n$, where 
$\gamma $ is the normal state specific heat coefficient, 
$T_{\rm c}$ the superconducting transition temperature, and 
$n$ the density of conduction charge carriers.
It includes contributions from the condensation energy $E_{\rm con}$, 
which is proportional to $\gamma T_{\rm c}^2$.
For example,
in a Gorter-Casimir two-fluid picture \cite{Gorter34}, using an
extended version \cite{Bardeen54} of the Ginzburg-Landau approximation \cite{Ginzburg50}
the corresponding contribution reads \cite{lip02}
\begin{equation}
e\slPhi_{\rm Therm} (\vec{r}) \approx 
\frac{n_s(\vec{r}) }{n} \;
\frac{\partial E_{\rm con}}{\partial n}
+ \sqrt{1-\frac{n_s(\vec{r})}{n}} \; \frac{T^2}{2} \; \frac{\partial \gamma }{\partial n}
\end{equation}
where $n_s(\vec{r})/n$ is the condensate fraction of the density. 
This term is referred to as the thermodynamic contribution to the 
electrostatic potential
\cite{Rickayzen69,Adkins68}.
A second contribution comes from the inertial forces and the Lorentz force
that act on the circulating supercurrent around the vortex
\cite{Bopp37,London50,Vijfeijken64}. 
This force decays on the scale of the London penetration depth, 
and is comparable in magnitude with the thermodynamic contribution near
the vortex core.
It arises due to the kinetic energy of the superflow, and 
also leads to a depletion of charge carriers in the vortex core region. 
In the inter-vortex regions, where the supercurrent momentum is 
determined by the phase gradient of the order parameter, 
$\grad \vartheta $, the corresponding contribution to the 
electrostatic potential reads \cite{Vijfeijken64}
\begin{equation}
e\slPhi_{\rm Bernoulli}(\vec{r} )\approx -\frac{n_s}{n} \; \frac{1}{2m^\ast } 
\left(\hbar \grad \vartheta (\vec{r}) -e^{\ast } \vec{A}(\vec{r})\right)^2
\end{equation}
where $e^\ast =2e$ and $m^\ast $ are the effective charge and mass of the Cooper pair,
and $\vec{A}$ is the vector potential.
The corresponding electrostatic potential is referred to as the 
Bernoulli potential.  The electrostatic potential resulting from both contributions 
leads to a charge depletion in the vortex core of order
\begin{equation}
\label{static}
\delta Q_{\rm static} \sim e \; \left( \frac{\slDelta}{E_f}\right)^2.
\end{equation}
In general, the electrostatic potential is determined by a screened Poisson 
equation \cite{Thomas}. However, the spatial variations of 
$n(\vec{r})$, $\slDelta (\vec{r})$, and $\vec{A}(\vec{r})$ occur on long-wavelength
superconducting scales, while screening takes place on the short-wavelength 
Thomas-Fermi screening length, $\lambda_{\rm TF}$. Thus, to
leading order in the quasiclassical expansion parameter, $\lambda_{\rm TF}/\xi_0$, 
the superconductor maintains local charge neutrality \cite{GK75,art79}.
The first corrections to the charge density are then given by
\begin{equation}
-\grad^2 \slPhi = 4\pi \rho(\vec{r}).
\end{equation}
with $\rho(\vec{r})=n(\vec{r})-n_0$, and $n_0$ is the charge density in
the normal state. We note that 
a possible charge pileup in a region of size $\lambda_{\rm TF}^2$ around the
vortex center {\it induced by} 
the presence of the superconducting state is expected to be small, because
the corresponding overlap region between such a charge
distribution and the superconducting order parameter is small by the
ratio $(\lambda_{\rm TF}/\xi_0)^2$ and in addition the order parameter
is small by a factor $\lambda_{\rm TF}/\xi_0$.\footnote{This 
last statement needs to
be modified if vortex core shrinking takes place at
low temperatures in superclean materials, however the general conclusions
will still hold.}
Consequently, any possible charge induced within the area $\lambda_{\rm TF}^2$
in the vortex center must be small by at least the ratio 
$(\lambda_{\rm TF}/\xi_0)^3$. This allows us to neglect such charges
and leads to an effective decoupling between charge variations on
the Thomas-Fermi screening length and the superconducting order parameter.
Charge fluctuation on the Thomas-Fermi length scale are
entirely determined by the normal state charge fluctuations, and the
leading order charge variations induced by the superconducting state are 
varying on the long-wavelength (superconducting) scales.

The entire picture discussed so far for the equilibrium vortex lattice
changes dramatically when considering a time-dependent
perturbation on a vortex. 
In order to reduce the Coulomb energy associated with the charge
accumulation an internal electro-chemical potential, $\slPhi(\vec{r};t)$,
develops in response to an external electric
field \cite{GK75}. 
This potential produces an internal electric field,
$\vec{E}^{\mbox{\tiny int}}(\vec{r};t)$, which is the same
order of magnitude as the external field. Even though the external field may
vary on a scale that is large compared to the coherence length, $\xi_0$,
the internal field
develops on the coherence length scale. The source of the internal field is a
charge density $\rho(\vec{r},t)$ that accumulates inhomogeneously over 
length scales of order
the coherence length. 
It is necessary to calculate the induced potential
self consistently from the spatially varying order parameter,
spectral function and distribution function for
the electronic states in the vicinity of the vortex core.
An order of magnitude estimate shows that to produce an induced
field of the order of the external field, the
dynamically induced charge is of order 
\begin{equation}
\label{dyn}
\delta Q_{\rm dynamic} \sim e\;\frac{\slDelta}{E_f}\;
\frac{\delta v_{\omega}}{\slDelta },
\end{equation}
where $\delta v_{\omega}\sim e{E}^{\mbox{\tiny ext}}\xi_0\slDelta/\omega$ is
the typical energy scale set by the strength of the external field.
This charge density accumulates predominantly in the vortex
core region and creates a dipolar field around the vortex core 
\cite{bar65,esc99}.
For a pinned vortex the charge accumulates near the
interface separating the metallic inclusion from the superconductor 
\cite{esc02}.
The charge in Eq.~\eqref{dyn} exceeds the 
dynamical charge that would result from
a dynamic motion of the vortex charges in Eq.~\eqref{static}, provided
$\delta v_{\omega} \gg \slDelta^2/E_f$. In this case, we can neglect the
contributions coming from the response of the charges that are present
already for a static vortex, and concentrate on the new contributions
Eq.~\eqref{dyn}.
Note that the dynamical charges are absent in the superclean case, where
the electric field is purely inductional \cite{kop94}.

In the next two sections we provide discussion of the 
expansion in small quasiclassical parameters, and
a summary of the
nonequilibrium quasiclassical equations, including the transport
equations for the quasiparticle distribution and spectral functions,
constitutive equations for the order parameter, impurity self-energy
and electromagnetic potentials. 
We then apply this theory to the response of a vortex in 
a chiral, spin-triplet superconductor to an {\em a.c.} electromagnetic
perturbation. The equilibrium properties of such a vortex have been
discussed in detail in an accompanying paper, Ref. \cite{Sauls09}.
In particular, we discuss the {\em a.c.} response of singly quantized and
doubly quantized vortices in Sec. \ref{sec-chiral_vortices}.

\section{Expansion in small parameters}

The physics of inhomogeneous metals and superconductors 
described by the quasiclassical approximation
is governed by well defined small expansion parameters. We will assume
that there is one such parameter that describes all small quantities 
in the system, and will assign to this parameter the notation $small $
\cite{ser83,rai95,rai94,esc99fluk}.
The typical microscopic
length scales of the problem,
in short denoted by $a_0$,
are the Bohr radius $a_B$, the Fermi wave
length $\lambda_f$, and  the Thomas-Fermi wave length $\lambda_{\rm TF}$.
The mesoscopic, superconducting length scales are the
coherence length $\xi_0$, and the penetration depth $\lambda $.
Correspondingly, large energy scales are the Fermi energy $E_f$, and
the Coulomb energy $e^2/a_0$, whereas small, quasiclassical energy
scales are the gap, $\slDelta $, the energy of the external perturbation, 
$\hbar \omega $,
and the energy related to the transition temperature, $k_{\rm B}T_{\rm c}$.
We assume that spatial variations near the vortex core are
on the scale $\xi_0$, and time variations on the scale $1/\omega $.
The normal state density of states at the Fermi level, $N_f$, is
of order $N_f\sim 1/E_fa_0^3$.
With this notation we have the following assignments that we need
to estimate the electromagnetic fields, charges and currents:
\begin{center}
\begin{tabular}{|c|c|c|}
\hline
$small^0$ & $small^1$ & $small^2 $\\
\hline 
\hline &&\\[-0.3cm]
$\displaystyle e^2N_f \sim \frac{1}{a_0^2} $ &  &\\[0.3cm]
\hline &&\\[-0.3cm]
$\displaystyle  \hbar v_f \sim \xi_0\slDelta  $ &
$\displaystyle \frac{v_f}{c} \sim \frac{a_0}{\lambda } $ & \\[0.3cm]
\hline &&\\[-0.3cm]
&$\displaystyle \hbar \vec{v}_f \cdot \grad \sim \slDelta $&
$\displaystyle \grad^2 \sim \frac{1}{\xi_0^2} $  \\[0.3cm]
\hline &&\\[-0.3cm]
&$\displaystyle \hbar \partial_t \sim \hbar \omega $&
$\displaystyle \frac{1}{c}\partial_t \sim \frac{a_0}{\lambda \xi_0} \frac{\hbar \omega }{\slDelta}$ \\[0.3cm]
\hline
\end{tabular}
\end{center}

\subsection{Electrostatic Fields and Potentials}

In equilibrium, the electrostatic fields are given in terms of
the potentials by $\vec{E}= -\grad \slPhi $, $\vec{B} = \grad \times \vec{A}$,
and fulfill the Maxwell's equations $\grad\cdot\vec{B}=0$, 
$\grad \times \vec{B} = \frac{4\pi}{c} \vec{j} $, $\grad\cdot \vec{E}= 4\pi \rho$,
$\grad \times \vec{E}=\vec{0}$.
The sources for the static fields in the superconductor are
the Meissner currents $\vec{j}$ and the small charges $\rho$ 
discussed in the introduction.
The Meissner current density in a superconductor can be estimated by noting
that
$\frac{e}{c}|\vec{j}| \sim e^2v_f N_f \slDelta /c \sim \slDelta/a_0\lambda 
\sim small^2$.
The electrostatic potentials induced in the superconducting state 
are of order $e\slPhi \sim \slDelta^2N_f \sim \slDelta^2 /E_f a_0^3$, and
estimating $E_f\sim v_fp_f/2 \sim \hbar v_f/\lambda_f \sim \slDelta \xi_0/a_0$
we arrive at $e\slPhi \sim a_0\slDelta/\xi_0 \sim small^2$.
The averaged static magnetic field $\bar B_0$ is given by the condition
that one flux quantum penetrates the vortex unit cell. In contrast, the
variation of the magnetic field on the coherence length scale, $\delta 
B_0$, is a factor $\xi_0/\lambda $ smaller than the averaged field.
This leads to the following estimates for physical parameters:
\begin{center}
\begin{tabular}{|c|c|c|c|}
\hline
$small $ & $small^2$ & $small^3 $& $small^4$\\
\hline 
\hline &&&\\[-0.3cm]
$\displaystyle e\bar B_0 \sim \frac{\slDelta}{a_0} $
&$\displaystyle \frac{e}{c} j \sim \frac{\slDelta}{a_0\lambda} $
&&
$\displaystyle e\rho \sim \frac{a_0\slDelta}{\xi_0^3} $\\[0.3cm]
\hline &&&\\[-0.3cm]
$ \displaystyle \frac{e}{c} v_f A \sim \slDelta $ & 
$\displaystyle e\slPhi \sim \frac{a_0\slDelta }{\xi_0} $&&\\[0.3cm]
\hline &&&\\[-0.3cm]
$\displaystyle e\delta B_0 \sim \frac{\xi_0}{\lambda a_0} \slDelta  $&&
$\displaystyle e E \sim \frac{a_0}{\xi_0^2}\slDelta $ & \\[0.3cm]
\hline 
\end{tabular}
\end{center}
In particular, for the free energy density the contributions from the 
Coulomb energy and the magnetic field energy are
\begin{equation}
\frac{\delta j \delta A}{2} \sim \slDelta^2 N_f\frac{\xi_0^2}{\lambda^2}  \sim small^2, \qquad
\frac{ \rho \slPhi}{2} \sim \slDelta^2 N_f \frac{a_0^4}{\xi_0^4}  \sim small^6\,.
\end{equation}
For the magnetic field energy the relevant quantity considered here 
is the contribution
that results from the deviations $\delta \vec{B}_0$ from the averaged field $\bar{\vec{B}}_0$
and its corresponding current $\delta \vec{j} = \frac{c}{4\pi } \grad \times \delta \vec{B}_0$.
Note that in the strong type II limit and for short coherence length 
superconductors (this is the case for example for cuprate superconductors,
where $\xi_0/\lambda\sim 0.01$ and $a_0/\xi_0\sim 0.2$), 
the scales can conspire such that 
$(\xi_0/\lambda)^2$ is of the same order as $(a_0/\xi_0)^4$.
In this case, the Coulomb energy contribution might become important for
the vortex lattice structure \cite{Bill}. 

\subsection{Electrodynamical Response}

For the dynamical response it is useful to split the vector potential
into transverse and longitudinal parts,
\begin{equation}
\vec{A}= \vec{A}_{\rm L} + \vec{A}_{\rm T}, \qquad \grad\cdot \vec{A}_{\rm T} =0, \qquad
\grad \times \vec{A}_{\rm L}=\vec{0}.
\end{equation}
The longitudinal part can be written as 
$\vec{A}_{\rm L}= \frac{c}{2e} \grad \zeta $, and can be gauged away by
changing the phase of the order parameter $\theta $ into 
$\theta + \zeta $ and the electrical potential $\slPhi $ into 
$\slPhi'= \slPhi + \frac{1}{2e} \partial_t \zeta $. Thus, we use a gauge
where $\vec{A}'=\vec{A}_{\rm T}$. 
The fields are given in terms of the potentials as
\begin{equation}
\vec{B}=\grad \times \vec{A}', \quad
\vec{E}= -\frac{1}{c} \partial_t \vec{A}' - \grad \slPhi' 
\,,
\end{equation}
and the Maxwell equations read (with the corresponding estimates of the various terms)
\begin{equation}
\label{estimate}
\underbrace{\frac{\;\; }{\;\; } \grad^2 \slPhi'}_{ small^3} = 4\pi \rho, \qquad
\underbrace{\frac{\;\; }{\;\; } \grad^2 \vec{A}'}_{ small^2}
+\underbrace{\frac{1}{c^2} \partial_t^2 \vec{A}'
+ \frac{1}{c} \partial_t \grad \slPhi'}_{small^4}
= \underbrace{\frac{4\pi }{c} \vec{j}}_{small^2 }.
\end{equation}

We drive the superconducting vortices out of equilibrium with an external
{\em  a.c. } electromagnetic field,
$\delta \vec{A}_\omega^{\rm ext } (t)= \delta \vec{A}_0 e^{-i\omega t }$,
of frequency $\omega \sim \slDelta /\hbar $.
The perturbing potential then is of the form
$\delta v^{\rm ext}_\omega = -\frac{e}{c} \vec{v}_f \cdot \delta \vec{A}^{\rm ext}_\omega (t)$. 
For the linear response approximation to hold, we assume that
$\delta v^{\rm ext}_\omega  \sim \hbar \omega \delta $ with $\delta  $ a 
small parameter that defines the region of applicability of linear response.
This means that for small frequencies
the vortex oscillation amplitude
will be $\sim \xi_0 \delta $, and the vortex velocity 
$v_{\rm L}\sim v_f (\hbar \omega /\slDelta ) \delta $.
We can then estimate the fields to be of order:
\begin{center}
\begin{tabular}{|c|c|c|c|}
\hline
$\delta $ & $small \cdot \delta $ & $small^2\cdot \delta $ & $small^3 \cdot  \delta $\\
\hline 
\hline &&&\\[-0.3cm]
$\displaystyle eA_\omega^{\rm ext} \sim \frac{\lambda}{a_0} \hbar \omega \delta$
&&$\displaystyle eE_\omega^{\rm ext} \sim \frac{(\hbar \omega )^2 }{\slDelta \xi_0} \delta $
& \\[0.3cm]
\hline &&&\\[-0.3cm]
$\displaystyle eA_\omega^{\rm int} \sim \frac{\xi_0^2}{\lambda a_0} \slDelta \delta$
&$ \displaystyle e\slPhi^{\rm int} \sim \hbar \omega \delta $ & 
&\\[0.3cm]
\hline &&&\\[-0.3cm]
&$\displaystyle eB_\omega^{\rm int} \sim \frac{\xi_0}{\lambda a_0} \slDelta \delta $
&$\displaystyle eE_\omega^{\rm int} \sim \frac{\hbar \omega \delta }{\xi_0}$
& $\displaystyle e|\grad \times \vec{E}_\omega^{\rm int}| \sim \frac{\hbar \omega \delta }{\lambda ^2}$ \\[0.3cm]
\hline &&&\\[-0.3cm]
&$\displaystyle e j_{\omega }\sim \frac{v_f }{a_0^2} \slDelta \delta $
&$\displaystyle \frac{e}{c} j_{\omega }\sim \frac{\slDelta \delta }{\lambda a_0} $
&$\displaystyle e\rho_{\omega } \sim \frac{\hbar \omega \delta }{\xi_0^2}$
\\[0.3cm]
\hline
\end{tabular}
\end{center}
Below, we use these estimates to simplify the coupled system of
Maxwell's equations and transport equations for the superconducting
charge carriers. In particular, we note that because
\begin{equation}
\grad\cdot \vec{j} \sim \frac{\slDelta^2 \delta }{e\hbar a_0^2} \sim small^2 \cdot \delta,
\qquad
\partial_t \rho \sim \frac{\slDelta^2\delta }{e\hbar \xi_0^2} \sim small^4 \cdot \delta ,
\end{equation}
the condition of local charge neutrality is fulfilled in 
linear response to within two orders in $small $.
Furthermore, the terms
\begin{equation}
\frac{1}{c^2}\partial_t^2 \vec{A}' + \frac{1}{c} \partial_t \grad \slPhi' \sim small^4 
\end{equation}
can be neglected compared to the terms
$\grad^2 \vec{A}' \sim small^2$ and
$ \frac{4\pi }{c} \vec{j} \sim small^2$.

\section{Nonequilibrium Transport Equations}

The quasiclassical theory describes equilibrium and nonequilibrium
properties of superconductors on length scales that are large compared
to microscopic scales (i.e. the lattice constant, Fermi wavelength,
$k_f^{-1}$, Thomas-Fermi screening length, $\lambda_{\rm TF}$, etc.) and
energies that are small compared to the atomic
scales (e.g. Fermi energy, $E_f$, plasma frequency,
conduction band width, etc.). Thus, the expansion parameter $small$ defines
the limits of validity of the quasiclassical
theory. In particular, we require $k_f\xi_0\gg 1$, $k_{\rm B} T_{\rm c}/E_f\ll 1$
and $\hbar\omega\ll E_f$, where the {\em a.c.} frequencies of interest
are typically of order $\slDelta/\hbar \sim k_{\rm B}T_{\rm c}/\hbar $, or smaller, and the
length scales of interest are of order the coherence length,
$\xi_0= \hbar v_f/2\pi k_{\rm B}T_{\rm c}$, or longer. Hereafter we use
units in which $\hbar=k_{\rm B}=1$, and adopt the sign convention
$e =-|e|$ for the electron charge.

In quasiclassical theory quasiparticle wavepackets move along nearly straight,
classical trajectories at the Fermi velocity. The classical dynamics of the
quasiparticle excitations is governed by semi-classical transport equations for
their phase-space distribution function.  The quantum mechanical degrees of
freedom are the ``spin'' and ``particle-hole degree
of freedom'', described by $4\times 4$ density matrices (Nambu matrices).
The quantum dynamics is coupled to the classical dynamics of the
quasiparticles in phase space through the matrix structure of the
quasiclassical transport equations.

The nonequilibrium quasiclassical transport equations
\cite{eil68,lar68,eli71,schmid75,LO75b}
are formulated in terms of a quasiclassical Nambu-Keldysh propagator
$\check{g}(\vec{p}_f,\vec{R};\epsilon,t)$,
which is a matrix in the combined Nambu-Keldysh space \cite{kel64}, 
and is a function of position $\vec{R}$,
time $t$, energy $\epsilon$, and  momenta $\vec{p}_f$ on the Fermi
surface. 
We denote Nambu-Keldysh matrices by a ``check'', and their
4$\times$4 Nambu submatrices by a ``hat''.  Thus, the Nambu-Keldysh
matrices for the quasiclassical propagator and self-energy have the form,
\be\label{Nambu-Keldysh}
\check{g}=\left(\begin{array}{cc}\hat{g}^{\rm R}&\hat{g}^{\rm K}\cr 0 &\hat{g}^{\rm A}\end{array}\right)
\qquad ,\qquad
\check{\sigma}=
\left(\begin{array}{cc}\hat{\sigma}^{\rm R}&\hat{\sigma}^{\rm K}\cr 0& \hat{\sigma}^{\rm A}\end{array}\right)
\,,
\ee
where $\hat{g}^{\rm R,A,K}$ are the retarded (R), advanced (A) and Keldysh (K)
quasiclassical propagators, and similarly for the self-energy functions.
Each of these components of $\check{g}$ and $\check{\sigma}$
are $4\times 4$ Nambu matrices in combined particle-hole-spin space.
For a review of the methods and an introduction to the notation we refer
to Refs. \cite{lar86,ser83,rai95}.
In  the compact Nambu-Keldysh notation the transport equations and  the
normalization conditions read
\begin{equation}\label{transpqcl}
\bigl[(\epsilon+{e\over c}\vec{v}_f\cdot\vec{A})\check\tau_3-eZ_0\slPhi \check{1}
-\check{\!\slDelta\,}_{\rm mf}-\check\nu_{\rm mf}-\check\sigma_{\rm i} \,,\check
g\bigr]_{\otimes} +i\vec{v}_f\cdot\grad \check g\\ = \ \check{0}\enspace ,
\end{equation}
\begin{equation}\label{normalize}
\check g\otimes \check g=-\pi^2\check{1}\enspace ,
\end{equation}
where the commutator is
$[\check A,\check B]_\otimes=\check A\otimes\check B-\check B\otimes\check A$,
\begin{equation}
\check A \otimes \check B(\epsilon,t) =
e^{\frac{i}{2}(\partial_\epsilon^A \partial_t^B-\partial_t^A\partial_\epsilon^B)}
\check A(\epsilon,t)\check B(\epsilon,t)
\,.
\end{equation}
The vector potential, $\vec{A}(\vec R;t)$, includes
$\vec{A}_0(\vec R)$ which generates the static magnetic field,
$\vec{B}_0 (\vec R) =\grad \times\vec{A}_0(\vec R )$, as well as
the non-stationary vector potential describing the time-varying
electromagnetic field; $\check{\!\slDelta\,}_{\rm mf}(\vec{p}_f,\vec{R};t)$
is the mean-field order parameter matrix,
$\check\nu_{\rm mf}(\vec{p}_f,\vec{R};t)$ describes diagonal
mean fields due to quasiparticle interactions (Landau interactions),
and $\check\sigma_{\rm i}(\vec{p}_f,\vec{R};\epsilon,t)$ is the impurity
self-energy. The electrochemical potential $\slPhi (\vec{R};t)$ includes
the field generated by the induced charge density, $\rho(\vec{R};t)$.
The coupling of quasiparticles to the external potential involves
virtual high-energy processes, which result from polarization of the
non-quasiparticle background. The interaction of quasiparticles with
both the external potential $\slPhi$ and the polarized background can be
described by coupling to an effective potential $Z_0\slPhi$ \cite{ser83}.
The high-energy renormalization factor $Z_0$ is defined below
in Eq. (\ref{renormalization}). The coupling of the quasiparticle
current to the vector potential in Eq. (\ref{transpqcl}) is given in
terms of the quasiparticle Fermi velocity. No additional renormalization
is needed to account for the effective coupling of the charge current to
the vector potential because the renormalization by the non-quasiparticle
background is accounted for by the effective potentials that
determine the band structure, and therefore the quasiparticle Fermi
velocity.

In quasiclassical theory the description in terms of
the variables ($\vec{p}_f,\vec{R};\epsilon $) is related to 
the ($\vec{p},\vec{R}$)
phase-space description by the transformation of distribution functions
$f(\vec{p}_f,\vec{R};\epsilon,t)$ to
$n(\vec{p},\vec{R};t)= f(\vec{p}_f,\vec{R};\epsilon(\vec{p},\vec{R};t),t)$, 
with $\epsilon(\vec{p},\vec{R};t)=\vec{v}_f(\vec{p}_f) \cdot (\vec{p}-\vec{p}_f) + 
\nu_{\rm mf} (\vec{p}_f, \vec{R};t) +
eZ_0\slPhi(\vec{R};t) -\frac{e}{c} \vec{v}_f(\vec{p}_f) \cdot \vec{A}(\vec{R};t) $,  see Ref. \cite{ser83}.

\subsection{Constitutive Equations}

Equations (\ref{transpqcl}-\ref{normalize})
must be supplemented by Maxwell's
equations for the electromagnetic potentials,
and by self-consistency equations
for the order parameter and the impurity self-energy.
We use the weak-coupling gap equation to describe
the superconducting state, including unconventional
pairing states. The mean field self energies are then given by,
\begin{equation}\label{gapequation}
\hat{\! \slDelta \,}^{\rm R,A}_{\rm mf}(\vec{p}_f,\vec{R};t)= N_f
\int_{-\epsilon_c}^{+\epsilon_c}{d\epsilon\over 4\pi i}
\big<
V(\vec{p}_f,\vec{p}_f^{\prime})
\hat f^{\rm K}(\vec{p}_f^{\prime},\vec{R};\epsilon,t)\big>
\enspace ,
\end{equation}
\begin{equation}\label{nuequation}
\hat\nu^{\rm R,A}_{\rm mf}(\vec{p}_f,\vec{R};t)= N_f
\int_{-\epsilon_c}^{+\epsilon_c}{d\epsilon\over 4\pi i}
\big<
A(\vec{p}_f,\vec{p}_f^{\prime})
\hat{\mbox{g}}^{\rm K}(\vec{p}_f^{\prime},\vec{R};\epsilon,t)\big>
\enspace ,
\end{equation}
\begin{equation}\label{Kequations}
\hat{\!\slDelta \,}^{\rm K}_{\rm mf}(\vec{p}_f,\vec{R};t)=0\enspace\,,
\qquad
\hat\nu^{\rm K}_{\rm mf}(\vec{p}_f,\vec{R};t)=0\enspace .
\end{equation}
The impurity self-energy,
\begin{equation}\label{born}
\check\sigma_{\rm i} (\vec{p}_f ,\vec{R};\epsilon,t)=
n_{\rm i} \; \check{t}(\vec{p}_f,\vec{p}_f,\vec{R};\epsilon, t)
\,,
\end{equation}
is specified by the impurity concentration, $n_{\rm i}$, and impurity
scattering $t$-matrix, which is obtained from the the self-consistent
solution of the $t$-matrix equations,
\ber\label{tmatrix}
\fl \check{t} (\vec{p}_f,\vec{p}''_f,\vec{R};\epsilon, t) =
\check{u}(\vec{p}_f,\vec{p}''_f)
+ N_f \big< \check{u}(\vec{p}_f,\vec{p}'_f)
\otimes \check{g}(\vec{p}'_f,\vec{R};\epsilon, t)
\otimes \check {t} (\vec{p}'_f,\vec{p}''_f,\vec{R};\epsilon, t) \big>
\,,
\eer
where the Fermi surface average is defined by
\begin{equation}\label{average}
\big<\ldots\big>\, =\
{1\over N_f}\int{d^2\vec{p}_f^{\prime}\over (2\pi)^3\mid\!\vec{v}_f^{\prime}\!\mid}\,
\left(\ldots\right)\,,\quad
N_f=\int{d^2\vec{p}_f^{\prime}\over (2\pi)^3\mid\vec{v}_f^{\prime}\!\mid} 
\,,
\end{equation}
and $N_f$ is the average density of states on the Fermi surface.
The Nambu matrix $\hat{f}^K$ ($\hat{\mbox{g}}^K$) is the off-diagonal (diagonal) 
part of $\hat g^K$ in particle-hole space. 
The other material parameters that enter the self-consistency equations are
the dimensionless pairing interaction, $N_fV(\vec{p}_f,\vec{p}_f^{\prime})$,
the dimensionless Landau interaction, 
$N_fA(\vec{p}_f,\vec{p}_f^{\prime})$,
the impurity concentration, $n_{\rm i}$, the impurity potential,
$\check u(\vec{p}_f,\vec{p}'_f)$, and the Fermi surface data:
$\vec{p}_f$ (Fermi surface), $\vec{v}_f(\vec{p}_f)$ (Fermi velocity).
We eliminate both the magnitude of the pairing
interaction and the cut-off, $\epsilon_c$,
in favor of the transition temperature, $T_{\rm c}$, using
the linearized, equilibrium form of the mean-field gap
equation (Eq. (\ref{gapequation})).

The quasiclassical equations are supplemented by constitutive
equations for the charge density, the current density and the induced
electromagnetic potentials. The formal result for the
nonequilibrium charge density, to linear order in $\slDelta/E_f$,
is given in terms of the Keldysh propagator by
\begin{equation}\label{density}
\hspace*{-2mm}
\fl \qquad \rho^{(1)}(\vec{R};t)=
eN_f\int_{-\epsilon_c}^{+\epsilon_c}{d\epsilon\over 4\pi i}
\big<Z(\vec{p}'_f)\,\mbox{Tr}\,
\left[\hat g^{\rm K}(\vec{p}_f^{\prime},\vec{R};\epsilon,t)\right]\big>
-2e^2N_fZ_0\slPhi(\vec{R};t)
\,,
\end{equation}
with the renormalization factors given by
\be\label{renormalization}
Z(\vec{p}_f)=1-\big\langle N_f A(\vec{p}'_f,\vec{p}_f )\big\rangle
\,,\quad
Z_0= \big\langle Z(\vec{p}_f) \big\rangle
\,.
\ee
The high-energy renormalization factor is related to an average
of the scattering amplitude on the Fermi surface by a Ward identity
that follows from the conservation law for charge \cite{ser83}.
The charge current induced by $\vec{A}(\vec{R};t)$, calculated to leading
order in $\slDelta/E_f $, is also obtained from the Keldysh
propagator,
\begin{equation}\label{current1}
\vec{j}^{(1)}(\vec{R};t)= eN_f
\int {d\epsilon\over
4\pi i}\mbox{Tr}\big<\vec{v}_f(\vec{p}_f^{\prime})\hat\tau_3
\hat g^{\rm K}(\vec{p}_f^{\prime},\vec{R};\epsilon,t) \big>
\,.
\end{equation}
There is no additional high-energy renormalization of the
coupling to the vector potential because the quasiparticle
Fermi velocity already includes the high-energy renormalization
of the charge-current coupling in Eq. \ref{current1}.
Furthermore, the {\it self-consistent} solution of the quasiclassical
equations for $\hat{g}^{\rm K}$ ensures the continuity equation for
charge conservation,
\begin{equation}
\label{conserv1}
\partial_t\;\rho^{(1)}(\vec{R};t)+\grad \cdot \vec{j}^{(1)}(\vec{R};t) = 0
\,,
\end{equation}
is satisfied.
An estimate of the contribution to the charge density
from the integral in Eq. (\ref{density})
leads to the condition of ``local charge neutrality'' \cite{GK75,art79}.
A charge density given by the elementary charge times the number of states
within an energy interval $\slDelta$ around the
Fermi surface implies $\rho^{(1)}\sim 2eN_f\slDelta $.
Such a charge density cannot be maintained within
a coherence volume because of the cost in Coulomb energy. This can
be seen in the estimate in Eq. \eqref{estimate}, where up to third
order in the parameter $small$ the charge fluctuations $\rho $ are suppressed.
This suppression, and the associated suppression in Coulomb energy, must
be taken care of by requiring the leading
order contribution to the charge density to vanish: i.e.
$\rho^{(1)}(\vec{R};t)=0$. Thus, the spatially varying
renormalized electro-chemical potential, $Z_0\slPhi$, is determined by
\be\label{localneutral}
2eZ_0\slPhi(\vec{R};t)=
\int_{-\epsilon_c}^{+\epsilon_c}{d\epsilon\over
4\pi i}\mbox{Tr}\big<
Z(\vec{p}'_f)
\hat g^{\rm K}(\vec{p}_f^{\prime},\vec{R};\epsilon,t)\big>
\,.
\ee
The continuity equation implies $\grad \cdot\vec{j}^{(1)}(\vec{R};t)=0$.
We discuss violations of the charge neutrality condition (\ref{localneutral}),
which are higher order in $\slDelta/E_f$, in Sec. \ref{chargeresp}.
Finally, Ampere's equation, with the current given by
Eq. (\ref{current1}), determines the
vector potential in the quasiclassical approximation,
\begin{equation}\label{vecpot}
\rot \rot \vec{A}(\vec{R};t)=
\frac{8\pi eN_f}{c}\int{d\epsilon\over 4\pi i}\mbox{Tr}\big<
\vec{v}_f(\vec{p}'_f) \hat\tau_3
\hat g^{\rm K}(\vec{p}_f^{\prime},\vec{R};\epsilon,t)\big>
\enspace .
\end{equation}

Equations (\ref{transpqcl})-(\ref{tmatrix}) and
(\ref{localneutral})-(\ref{vecpot}) constitute
a complete set of equations for calculating the
electromagnetic response of vortices in the quasiclassical
limit. For high-$\kappa$ superconductors we can simplify
the self-consistency calculations to some degree.
Since quasiparticles couple to the vector potential via
$\frac{e}{c} \vec{v}_f \cdot \vec{A}$, Eq. (\ref{vecpot}) shows that
this quantity is of order $8\pi e^2 N_fv_f^2/c^2=1/\lambda^2$,
where $\lambda$ is the magnetic penetration depth.
Thus, for $\kappa=\lambda/\xi_0\gg 1$, as in the layered cuprates,
the feedback effect of the current density on the vector potential
is small by factor $1/\kappa^2$.

\subsection{Linear Response}

For sufficiently weak fields we can calculate the electromagnetic response
to linear order in the external field. The propagator and the
self-energies are separated into unperturbed equilibrium parts
and terms that are first-order
in the perturbation,
\begin{equation}\label{linresp}
\check{g}=\check{g}_0+\delta\check{g}\,, \,\,
\check{\!\slDelta \,}_{\rm mf}=\check{\!\slDelta \,}_{0} +\delta\check{\!\slDelta \,}_{\rm mf}\,, \,\,
\check{\sigma}_{\rm i}=\check{\sigma}_{0}+\delta\check{\sigma}_{\rm i}
\enspace ,
\end{equation}
and similarly for the electromagnetic potentials,
$\vec{A}=\vec{A}_0+\delta\vec{A}$, $\slPhi=\delta\slPhi$.
The equilibrium propagators obey the matrix transport equation,
\begin{equation}\label{transpqcl0}
\bigl[(\epsilon+{e\over c}\vec{v}_f\cdot\vec{A}_0)
\check\tau_3-\check{\! \slDelta\,}_{0}-\check\sigma_{0}\,,\check
g_0\bigr] +i\vec{v}_f\cdot\grad \check g_0  = \ \check{0}
\,.
\end{equation}
These equations are supplemented by the
self-consistency equations for the mean fields,
Eqs. (\ref{gapequation})-(\ref{nuequation}),
the impurity self energy, Eqs. (\ref{born})-(\ref{tmatrix}),
the local charge-neutrality condition for the scalar potential,
Eq. (\ref{localneutral}), Amp\`ere's equation for the vector
potential, Eq. (\ref{vecpot}),
the equilibrium normalization conditions,
\begin{equation}\label{normalize0}
\check{g}_0^2=-\pi^2\check{1}
\,,
\end{equation}
and the equilibrium relation between the Keldysh function
and equilibrium spectral density,
\begin{equation}\label{KMS}
\hat{g}^{\rm K}_0=\tanh\left(\frac{\epsilon}{2T}\right)
\Big[\hat{g}^{\rm R}_0-\hat{g}^{\rm A}_0\Big]
\,.
\end{equation}

The first-order correction to the matrix propagator
obeys the linearized transport equation,
\begin{equation}\label{transpqcl1}
\fl \qquad \bigl[(\epsilon+{e\over c}\vec{v}_f\cdot\vec{A}_0)
\check\tau_3-\check{\!\slDelta \,}_{0}-\check\sigma_{0}\,,\delta\check
g\bigr]_{\otimes} +i\vec{v}_f\cdot\grad \delta\check g\\ = \
\bigl[
\delta\check{\!\slDelta \,}_{\rm mf}+\delta\check{\sigma}_{\rm i}+\delta\check{v}\,,\
\check{g}_0]_{\otimes}\enspace ,
\end{equation}
with source terms on the right-hand side from both
the external field ($\delta\check{v}$) and the internal
fields ($\delta\check{\!\slDelta \,}_{\rm mf}$, $\delta\check{\sigma}_{\rm i}$). 
In addition,
the first-order propagator satisfies the ``orthogonality condition'',
\begin{equation}\label{normalize1}
\check{g}_0\otimes\delta\check{g}
+\delta\check{g}\otimes\check{g}_0=\check{0}\enspace .
\end{equation}
obtained from linearizing the full normalization condition.
Note that the convolution product between an equilibrium and a
nonequilibrium quantity simplifies after Fourier transforming
$t\rightarrow\omega$:
\begin{equation}
\fl \qquad
\check A_0 \otimes \check B (\epsilon ,\omega )
= \check A_0 (\epsilon + \omega/2 )\check B (\epsilon ,\omega ), \qquad
\check B (\epsilon ,\omega ) \otimes \check A_0
= \check B (\epsilon ,\omega ) \check A_0 (\epsilon - \omega/2 ).
\end{equation}
The system of linear equations are supplemented
by the equilibrium and first-order self-consistency conditions for the
order parameter,
\begin{equation}\label{lgapequation0}
\hat{\! \slDelta \,}^{\rm R,A}_{0}(\vec{p}_f,\vec{R})=
N_f\int_{-\epsilon_c}^{+\epsilon_c}{d\epsilon\over 4\pi i}
\big< V(\vec{p}_f,\vec{p}'_f)
\hat f^{\rm K}_0(\vec{p}_f^{\prime},\vec{R};\epsilon)\big>
\,,
\end{equation}
\begin{equation}\label{lgapequation1}
\delta\hat{\!\slDelta \,}^{\rm R,A}_{\rm mf}(\vec{p}_f,\vec{R};t)=
N_f\int_{-\epsilon_c}^{+\epsilon_c}{d\epsilon\over 4\pi i}
\big< V(\vec{p}_f,\vec{p}'_f)
\delta\hat f^{\rm K}(\vec{p}_f^{\prime},\vec{R};\epsilon,t)\big>,
\end{equation}
and the impurity self-energy,
\ber\label{respborn0}
\hspace*{-5mm}
\fl \qquad
\check\sigma_{0} (\vec{p}_f ,\vec{R};\epsilon) &=&
n_{\rm i}\;\check{t}_0(\vec{p}_f,\vec{p}_f,\vec{R};\epsilon )\,,
\\
\fl \qquad
\hspace*{-5mm}
\label{tmatrix0}
\check{t}_0 (\vec{p}_f,\vec{p}''_f,\vec{R};\epsilon )&=&
\check {u} (\vec{p}_f,\vec{p}''_f)+
N_f \big< \check {u} (\vec{p}_f,\vec{p}'_f)
\check{g}_0(\vec{p}'_f,\vec{R};\epsilon )
\check {t}_0 (\vec{p}'_f,\vec{p}''_f,\vec{R};\epsilon ) \big>\,,
\\
\fl \qquad
\hspace*{-5mm}
\label{tmatrix1}
\delta \check\sigma_{\rm i} (\vec{p}_f ,\vec{R};\epsilon,t)&=&
n_{\rm i} N_f \big< \check {t}_0 (\vec{p}_f,\vec{p}'_f,\vec{R};\epsilon )
\otimes \delta \check{g}(\vec{p}'_f,\vec{R};\epsilon, t)
\otimes \check {t}_0 (\vec{p}'_f,\vec{p}_f,\vec{R};\epsilon ) \big>
\,.
\eer
The Keldysh matrix components of the last equation read explicitly 
\ber
\fl \delta \hat{\sigma}_{\rm i}^{\rm R,A} (\vec{p}_f ,\vec{R};\epsilon,t)&=&
n_{\rm i} N_f \big< \hat{t}_0^{\rm R,A} (\vec{p}_f,\vec{p}'_f,\vec{R};\epsilon )
\otimes \delta \hat{g}^{\rm R,A}(\vec{p}'_f,\vec{R};\epsilon, t)
\otimes \hat{t}_0^{\rm R,A} (\vec{p}'_f,\vec{p}_f,\vec{R};\epsilon ) \big>
\,,\\
\fl \delta \hat{\sigma}_{\rm i}^{\rm a} (\vec{p}_f ,\vec{R};\epsilon,t)&=&
n_{\rm i} N_f \big< \hat{t}_0^{\rm R} (\vec{p}_f,\vec{p}'_f,\vec{R};\epsilon )
\otimes \delta \hat{g}^{\rm a}(\vec{p}'_f,\vec{R};\epsilon, t)
\otimes \hat{t}_0^{\rm A} (\vec{p}'_f,\vec{p}_f,\vec{R};\epsilon ) \big>
\,,
\eer
with the `anomalous' propagator and self energy
\ber
\delta \hat{\sigma}^{\rm a} &=&
\delta \hat{\sigma}^{\rm K}-
\delta \hat{\sigma}^{\rm R}
\tanh\left(\frac{\epsilon-\omega/2}{2T}\right) 
+
\tanh\left(\frac{\epsilon+\omega/2}{2T}\right) 
\delta \hat{\sigma}^{\rm A}, \\
\delta \hat{g}^{\rm a} &=&
\delta \hat{g}^{\rm K} -
\delta \hat{g}^{\rm R}
\tanh\left(\frac{\epsilon-\omega/2}{2T}\right) 
+
\tanh\left(\frac{\epsilon+\omega/2}{2T}\right) 
\delta \hat{g}^{\rm A} .
\eer

In general the diagonal mean fields also contribute to the response.
However, we do not expect Landau interactions to lead to qualitatively
new phenomena for the vortex dynamics, so we have neglected these
interactions in the following analysis
and set $A(\vec{p}_f,\vec{p}_f')=0$ (i.e. $\check \nu_{\rm mf}=0$).
As a result the local charge neutrality condition for the
electro-chemical potential becomes,
\be\label{resplocalneutral}
2e\delta\slPhi(\vec{R};t)=\int_{-\epsilon_c}^{+\epsilon_c}{d\epsilon\over
4\pi i}\mbox{Tr}\big< \delta \hat g^{\rm K}(\vec{p}_f,\vec{R};\epsilon,t)\big>
\,.
\ee

In what follows we work in a gauge in which the induced
electric field, $\vec{E}^{\mbox{\tiny ind}}(\vec{R};t)$,
is obtained from $\delta\slPhi(\vec{R};t)$ and the uniform
external electric field, $\vec{E}_{\omega}^{\mbox{\tiny ext}}(t)$,
is determined by the vector potential $\delta\vec{A}_\omega (t)$.
For $\lambda/\xi_0\gg 1$ we can safely neglect corrections
to the vector potential due to the induced current.
Thus, in the Nambu-Keldysh matrix notation the electromagnetic
coupling to the quasiparticles is given by
\begin{equation}\label{perturbation}
\delta\check{v} =
-{e\over c}\vec{v}_f\cdot\delta\vec{A}_\omega (t)\check{\tau}_3
+e\delta\slPhi(\vec{R};t)\check{1}
\enspace .
\end{equation}
The validity of linear response theory requires the external
perturbation $\delta \check{v}$ be sufficiently small and that
the induced vortex motion responds to the external field at
the frequency set by the external field. At very low frequencies
frictional damping of the vortex motion, arising from the finite mean
free path of quasiparticles scattering from impurities, gives rise to a
nonlinear regime in the dynamical response of a vortex.
This regime is discussed extensively in the literature \cite{LO75b},
and is not subject of our study. However, for sufficiently small
field strengths the vortex motion is non-stationary over any time interval,
although it may be regarded as quasi-stationary at low enough frequencies.
The non-stationary motion of the vortex can be described by
linear response theory if $\delta\check{v}\ll 1/\tau$ for
$\omega\lsim 1/\tau$, and $\delta \check{v}\ll \omega$ for
$\omega\gsim 1/\tau$. Note that the frequency of the perturbation,
$\omega$, is not required to be small compared to the gap
frequency; it is only restricted to be small compared to
atomic scale frequencies, e.g $\omega \ll E_f/\hbar$.

Self-consistent solutions of Eqs. (\ref{lgapequation1}), (\ref{tmatrix1})
and (\ref{resplocalneutral}) for the self-energies and scalar potential
are  fundamental to obtaining a physically sensible solution for the
electromagnetic response. The dynamical self-energy corrections
are equivalent to ``vertex corrections'' in the Kubo formulation of linear
response theory. They are particularly important in the context of
nonequilibrium phenomena in inhomogeneous
superconductors. Vertex corrections usually vanish in
homogeneous superconductors because of translational and rotational symmetries.
Inhomogeneous states break these symmetries and
typically generate non-vanishing vertex corrections.
In our case these corrections are of vital importance;
the self-consistency conditions enforce charge conservation.
In particular, Eqs. (\ref{respborn0})-(\ref{tmatrix1}) imply charge
conservation in scattering processes,  whereas (\ref{lgapequation0})
and  (\ref{lgapequation1}) imply charge conservation in particle-hole
conversion processes; any charge which is lost (gained)
in a particle-hole conversion process is compensated by a corresponding
gain (loss) of condensate charge. It is the coupled quasiparticle and
condensate dynamics which conserves charge in superconductors.
Neglecting the dynamics of either component, or using a
non-conserving approximation for the coupling leads to unphysical results.
Self-consistent calculations
for the local excitation spectrum (spectral density)
also provides
key information for the interpretation of the dynamical response.
Because of particle-hole coherence
the spectral density is sensitive to the phase winding
and symmetry of the order parameter, as well as material properties
such as the transport mean-free path and impurity cross-section.

In the limit $\omega \to 0$ the equations above have to be modified,
in order to eliminate the zero modes associated with the stationary 
motion of the vortex lattice. However, it is possible also to
obtain solutions for this limit in quasiclassical theory.
For completeness, we provide the expression for the 
transport current density that
arises in the low-frequency limit, in the flux flow regime \cite{lar86,gor73,kop95}:
\begin{eqnarray}
\fl \qquad \frac{1}{c}\vec{j}_{\rm tr} \times \vec{\Phi}_0 = \nonumber \\
\fl \qquad -N_f\int \limits_{\rm cell} d^2\vec{R} \int \frac{d\epsilon }{8\pi i} \mbox{Tr} \left\langle
\left[ \left(\grad - 2i\frac{e}{c} \vec{v}_f \cdot \vec{A}_0 \hat \tau_3 \right)
\hat{\!\slDelta \,}_{0} + \frac{e}{c} \vec{v}_f \times \left( \grad \times \vec{A}_0 \right) \hat \tau_3 \right]
\delta \hat g^{\rm nst} \right\rangle
\end{eqnarray}
with $\delta \hat g^{\rm nst} = \lim_{\omega \to 0} \left[ \delta g^{\rm K} -
\tanh(\epsilon/2T) (\delta \hat g^{\rm R} - \delta \hat g^{\rm A})\right]$, 
that is obtained from a systematic 
expansion for small $\omega $.
The leading terms are proportional to the vortex velocity $v_{\rm L}$.

\section{Nonequilibrium Response of chiral vortices}\label{sec-chiral_vortices}

The dynamics of the electronic excitations of the
vortex core plays a key role in the dissipative processes in
type II superconductors. Except in the dirty limit, $\ell \ll \xi_0$,
the response of the core states to an electromagnetic field is
generally very different from that of normal electrons.
It is energetically unfavorable to maintain a
charge density of the order of an elementary charge over
a region with diameter of order of the coherence length.
Instead, an electrochemical potential is induced which
ensures that almost no net charge accumulates in the core
region. On the other hand, a dipolar-like charge distribution develops
which generates an internal electric field in the core.
The internal field varies on the scale of the coherence
length. This leads to a nonlocal response of the quasiparticles
to the total electric field, even when the applied field varies
on a much longer length scale and can be considered homogeneous.
The dynamical response of the vortex core includes the collective
mode of the inhomogeneous order parameter.
This mode couples to the electro-chemical
potential, $\delta\slPhi$, in the vortex core region.
This potential is generated by
the charge dynamics of vortex core states and gives rise to
internal electric fields which in turn drive the current density
and the order parameter near the vortex core region.
The induced electric fields in the core are the same order
of magnitude as the external field. The dynamics of the
core states are strongly coupled to the charge
current and collective mode
of the order parameter. Thus, the determination of the induced
order parameter, as well as the spectrum and distribution
function for the core states and nonequilibrium impurity scattering
processes requires dynamical self consistency.
Numerous calculations of the {\rm a.c.} response neglect the self-consistent
coupling of the collective mode and the spectral dynamics, or concentrate
on the $\omega\rightarrow 0$ limit \cite{kop78a,jan92,hsu95,kop95}.
Presently, quasiclassical theory is the only formulation of the theory of
nonequilibrium superconductivity
capable of describing the nonlocal response of the order parameter
and quasiparticle dynamics in the presence of mesoscopic
inhomogeneities and disorder.
The numerical solution to the self-consistency problem was presented for
unpinned $s$-wave and $d$-wave  vortices in Ref. \cite{esc97,esc99,esc01} and for pinned 
$s$-wave  vortices in Ref. \cite{esc02}.
In the following we present our results for a vortex in a chiral, spin-triplet
superconductor.
As discussed in our accompanying paper \cite{Sauls09}, the order parameter we
consider is of the form
\be\label{local} 
\vec{\vDelta}(\hat{\vec{p}},\vec{R})
= 
\vec{d}\,  
\Bigg[ 
|\slDelta_{+}(\vec{R})|\,e^{i m\phi}\,\frac{\left(\hat{\vec{p}}_x + i\hat{\vec{p}}_y\right)}{\sqrt{2}}
+
|\slDelta_{-}(\vec{R})|\,e^{i p\phi}\,\frac{\left(\hat{\vec{p}}_x - 
i\hat{\vec{p}}_y\right)}{\sqrt{2}}
\Bigg] 
\,.  
\ee            
We have shown, that stable vortex structures can occur for $m=-1, p=1$
(singly quantized vortex), and for $m=-2,p=0$ (doubly quantized vortex with
nearly homogeneous core).
We will in the following discuss for these two vortex structures the dynamical charge
response, the induced current density, and the magnetic field response.
Our calculations were performed with impurity scattering included in Born scattering limit for a mean free path of $\ell =10 \xi_0$, where $\xi_0=\hbar v_f/2\pi k_B T_c$ is the coherence length. The Fermi surface parameters are assumed to be isotropic, and the temperature was chosen to be $T=0.2\, T_c$. For simplicity we also assumed the high-$\kappa $ limit, where the penetration depth is large compared to the coherence length. 
Our calculations of the vortex structure are appropriate to the low-field limit near $H_{c1}$ where vortices are well separated from each other. 
The order parameter, impurity self energy, and the equilibrium spectra and current densities were obtained self consistently using the Riccati formulation of the quasiclassical transport theory with impurity and pairing self-energies
\cite{esc00}.

\subsection{Dynamical charge response}
\label{chargeresp}

The charge dynamics of layered superconductors has two distinct origins.
The c-axis dynamics is determined by the Josephson coupling between
the conducting planes. Here we are concerned with the in-plane
electrodynamics associated with the response of the order parameter
and quasiparticle states bound to the vortex core. We assume strong
Josephson coupling between different layers, and neglect variations
of the response between different layers. This requires that the
polarization of the electric field be in-plane, so that there is no
coupling of the in-plane dynamics to the Josephson plasma modes.
The external electromagnetic field is assumed to be long wavelength
compared to the size of the vortex core, $\lambda_{\mbox{\tiny EM}}\gg\xi_0$.
In this limit we can assume the {\em a.c.} electric field to be uniform and described
by a vector potential, $\vec{E}_{\omega}(t)=-\frac{1}{c}\partial_t\vec{A}_{\omega}$.
We can also neglect the response to the {\em a.c.} magnetic field
in the limit $\lambda\gg\xi_0$. In this case the spatial variation of the
induced electric field occurs mainly within each conducting layer
on the scale of the coherence length, $\xi_0$. Poisson's equation
implies that induced charge densities are of order $\delta\slPhi/\xi_0^2$,
where $\delta\slPhi$ is the induced electrochemical potential in the core.
This leads to a dynamical charge of order $e\,(\slDelta/E_f)$ in the vortex core.
Once the electrochemical potential is calculated from
Eq. (\ref{resplocalneutral}) we can calculate the charge density
fluctuations of order $(\slDelta/E_f)^3$ from Poisson's equation,
\be
\rho^{(3)}(\vec{R};t)=-\frac{1}{4\pi}\grad^2\,\delta \slPhi(\vec{R};t)
\,.
\ee

\begin{figure}[h]
%\begin{tabular}{c}
\centerline{
\begin{minipage}{0.36\hsize}
\epsfxsize0.7\hsize\epsffile{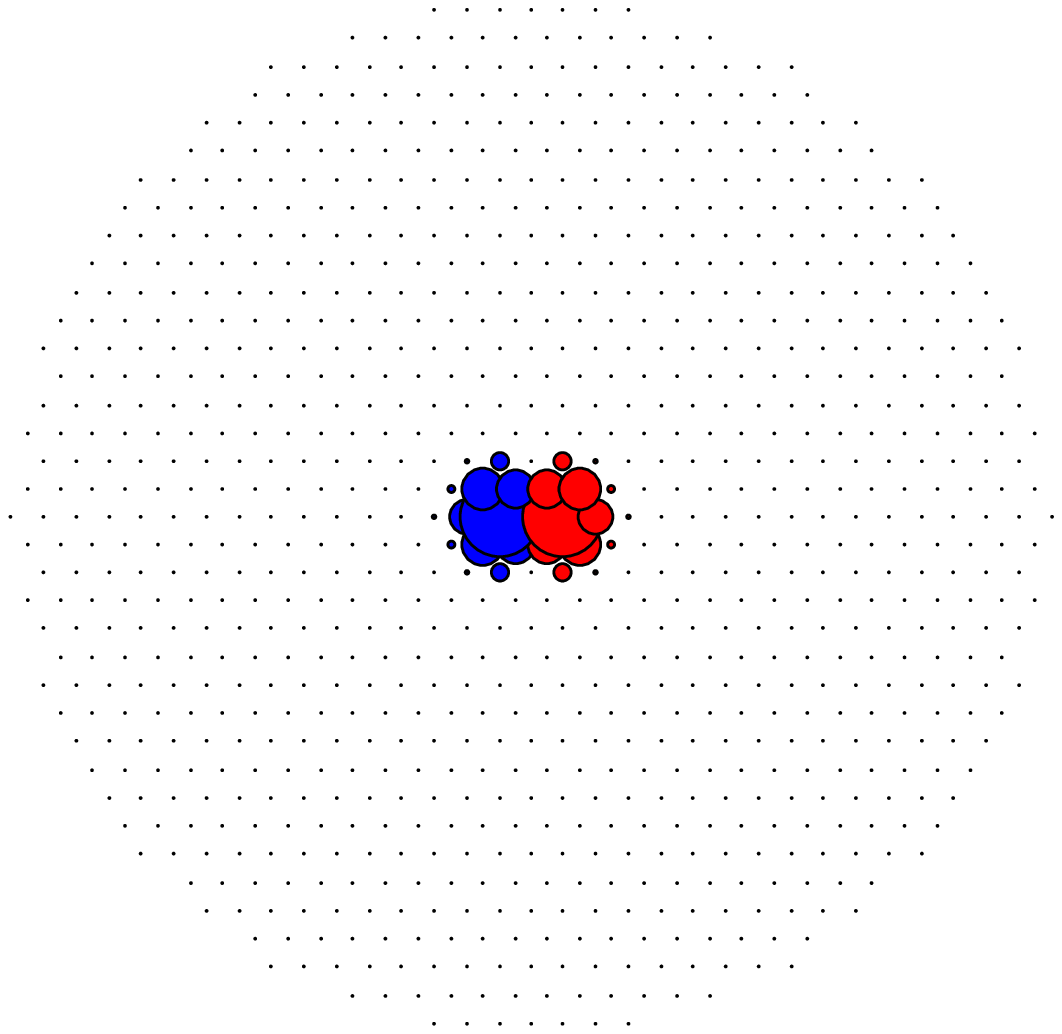}
\end{minipage}
\begin{minipage}{0.36\hsize}
\epsfxsize0.7\hsize\epsffile{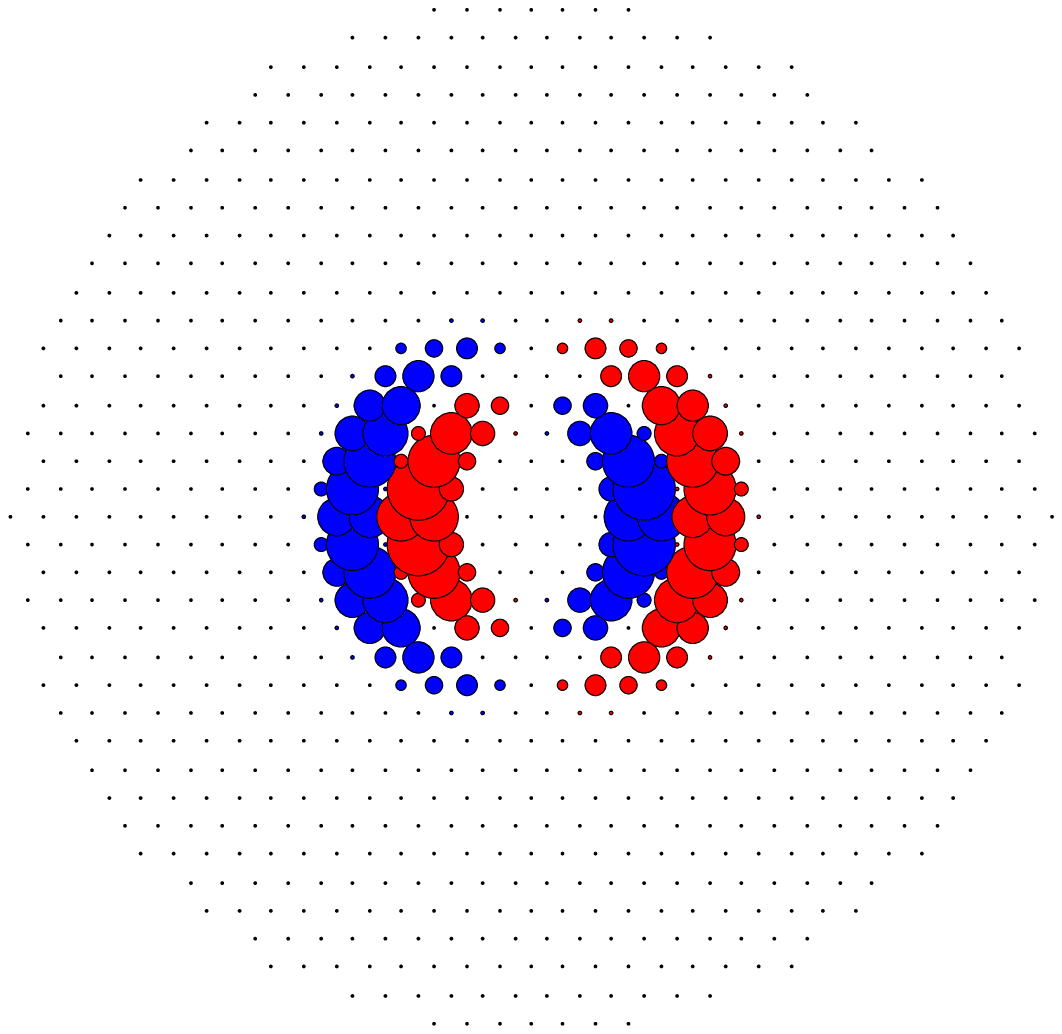}
\end{minipage}
}
%\end{tabular}
\caption{ 
\label{Fig1}
Out of phase charge response 
for the case of an $m=-1, p=1$ vortex (left) and for
the case of an $m=-2, p=0$ vortex (right) 
to an {\em a.c.} electric field with $\omega = 0.2 \slDelta$
polarized in $x$-direction.
Red denotes negative values for the response, blue positive.
}
\end{figure}

Fig. \ref{Fig1} shows
the charge distribution for $\omega=0.2\slDelta$ which oscillates out of phase
with the external field. We consider two vortex structures.
On the left hand side we show results for a singly quantized vortex 
with quantum numbers $m=-1, p=1$, and on the right hand side the corresponding
results for the doubly quantized vortex with quantum numbers $m=-2,p=0$.
For the singly quantized vortex we observe a charge dipole that oscillates with
the external frequency. The dipole vector is parallel  to the external 
electric field and the charge accumulation concentrated to the vortex core
area. For the doubly quantized vortex, shown on the right hand side in 
Fig.~\ref{Fig1}, a very different picture emerges.
A dipolar charge distribution accumulates
at the interface between the dominant and the subdominant order parameter
components,
oscillating at the frequency of the external field.
The induced charge which accumulates is of order of $e\slDelta/E_f$ within
a region of order $\xi_0^2$ in each conducting layer.
As discussed in the introduction, this charge is a factor of
$(\delta v_\omega E_f)/\slDelta^2$ larger than the 
static charge of a vortex that arises from
particle-hole asymmetry \cite{kho95,fei95,bla96}.

\subsection{Induced current density}

\begin{figure}[h]
%\begin{tabular}{c}
\centerline{
\begin{minipage}{0.36\hsize}
\epsfxsize0.7\hsize\epsffile{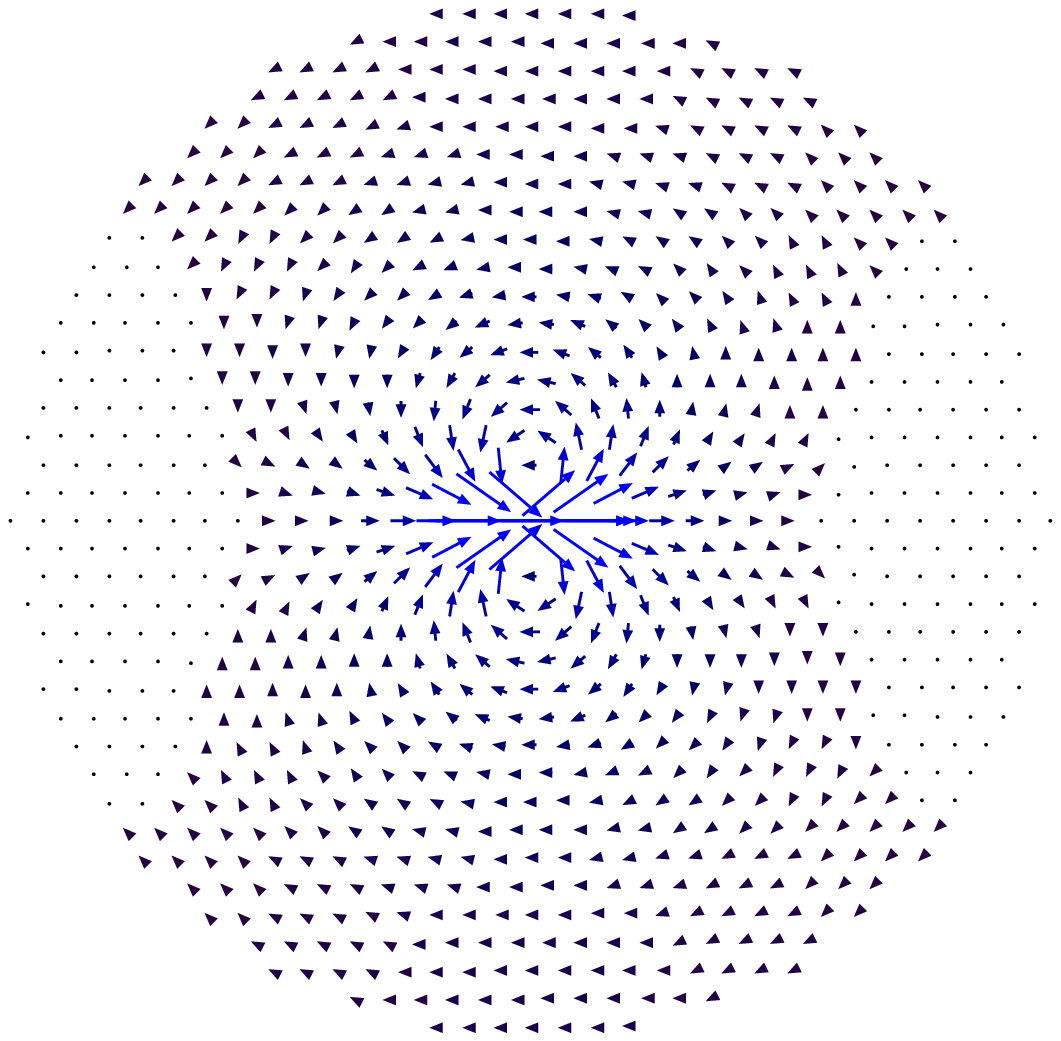}
\end{minipage}
\begin{minipage}{0.36\hsize}
\epsfxsize0.7\hsize\epsffile{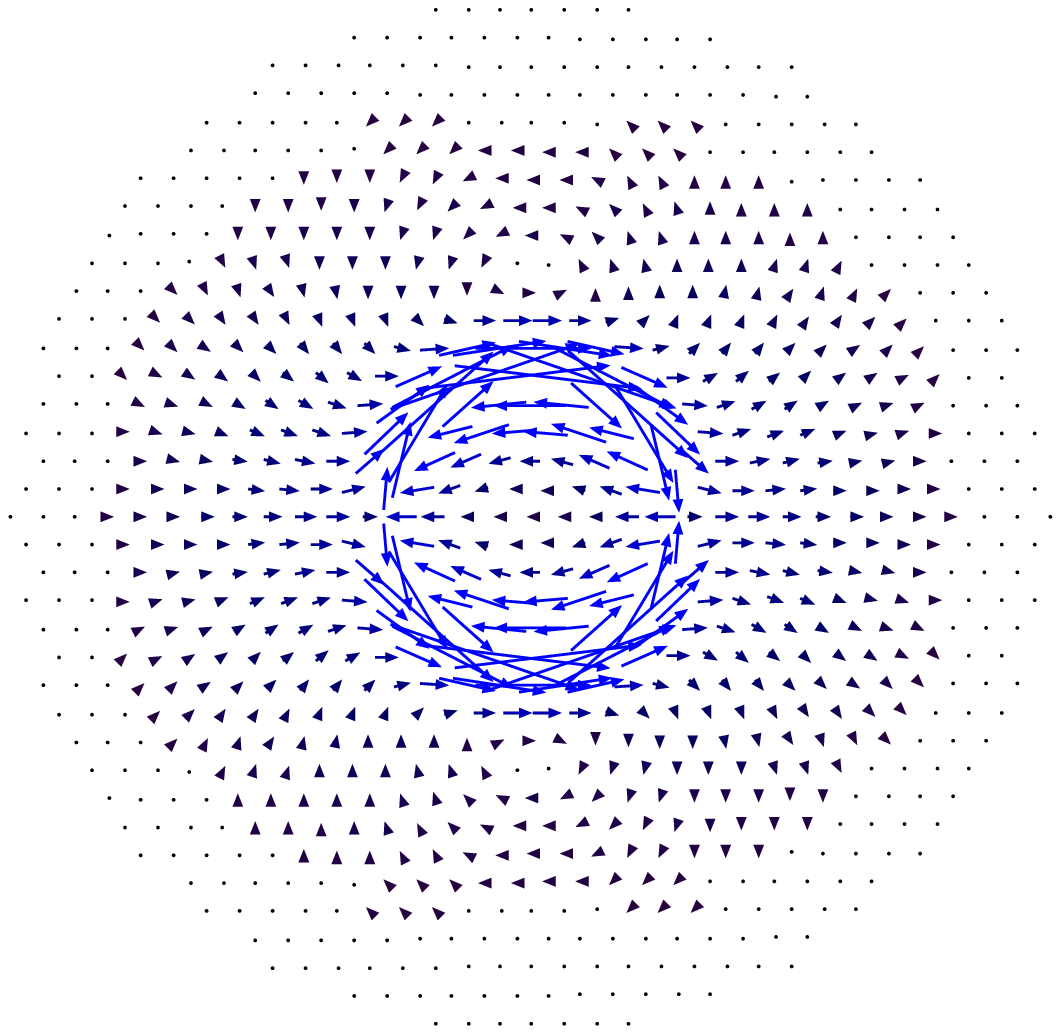}
\end{minipage}
}
%\end{tabular}
\caption{ 
\label{Fig2}
Absorptive current response
for the case of an $m=-1, p=1$ vortex (left) and for
the case of an $m=-2, p=0$ vortex (right) 
to an {\em a.c.} electric field with $\omega = 0.2 \slDelta$
polarized in $x$-direction.
}
\end{figure}

In the static vortex structure a circulating supercurrent is present
due to the screening of the quantized magnetic field penetrating the vortex.
This equilibrium current has been calculated in Ref.~\cite{Sauls09}.
For a doubly quantized vortex the circulating current of the vortex has a 
non-trivial structure, with currents in the vortex core that flow counter to 
the circulating supercurrent at large distances from the vortex center. 
Here, we discuss the additional, dynamical part of the current that is 
induced in the presence of the external {\em a.c.} electromagnetic field,
and
concentrate on the absorptive component of the current response, which is 
{\it in phase} with the external field. Results for the {\em a.c.} component of the current density 
are shown in Fig. \ref{Fig2} for $\omega=0.2\slDelta$.
The in-phase current response for a singly quantized vortex, shown
on the left hand side, shows a dipolar pattern, indicating
a region of strong absorption ($\delta \vec{j}||\delta \vec{E}^{\mbox{\tiny ext}}$) in the
vortex core, and emission
($\delta \vec{j}\cdot\delta \vec{E}^{\mbox{\tiny ext}}<0$) in the region roughly perpendicular
to the direction of the applied field several coherence lengths away from
the vortex center. 
The picture here is similar to that for an $s$-wave vortex in an external
{\em a.c.} field, discussed in our previous work \cite{esc97,esc99,esc01,esc02}.
Energy absorbed in the core is transported away from the vortex
center by the vortex core excitations in directions predominantly
perpendicular to the applied field. The net absorption is ultimately
determined by inelastic scattering and requires integrating the local
absorption and emission rate over the vortex array.
Note that the long-range dipolar component
does not contribute to the {\it total} dissipation.
Far from the vortex core the current response is
out of phase with the electric field and
predominantly a non-dissipative supercurrent.

For the doubly quantized vortex, the  response of which is shown on the
right hand side in Fig.~\ref{Fig2}, we find that the main absorption
results from the regions where the two time reversed order parameter phases
are overlapping. The current response here is strongly linked to the
order parameter response. Absorptive currents are flowing predominantly in the
region of the `domain wall' between the two components. Note that the
leading contribution in a multipole expansion with respect to the 
vortex center is again the dipolar term, although the pattern here has 
strong quadrupolar and higher order contributions.

\subsection{Magnetic field response}

\begin{figure}[h]
\centerline{
%\begin{tabular}{c}
\begin{minipage}{0.36\hsize}
\epsfxsize0.7\hsize\epsffile{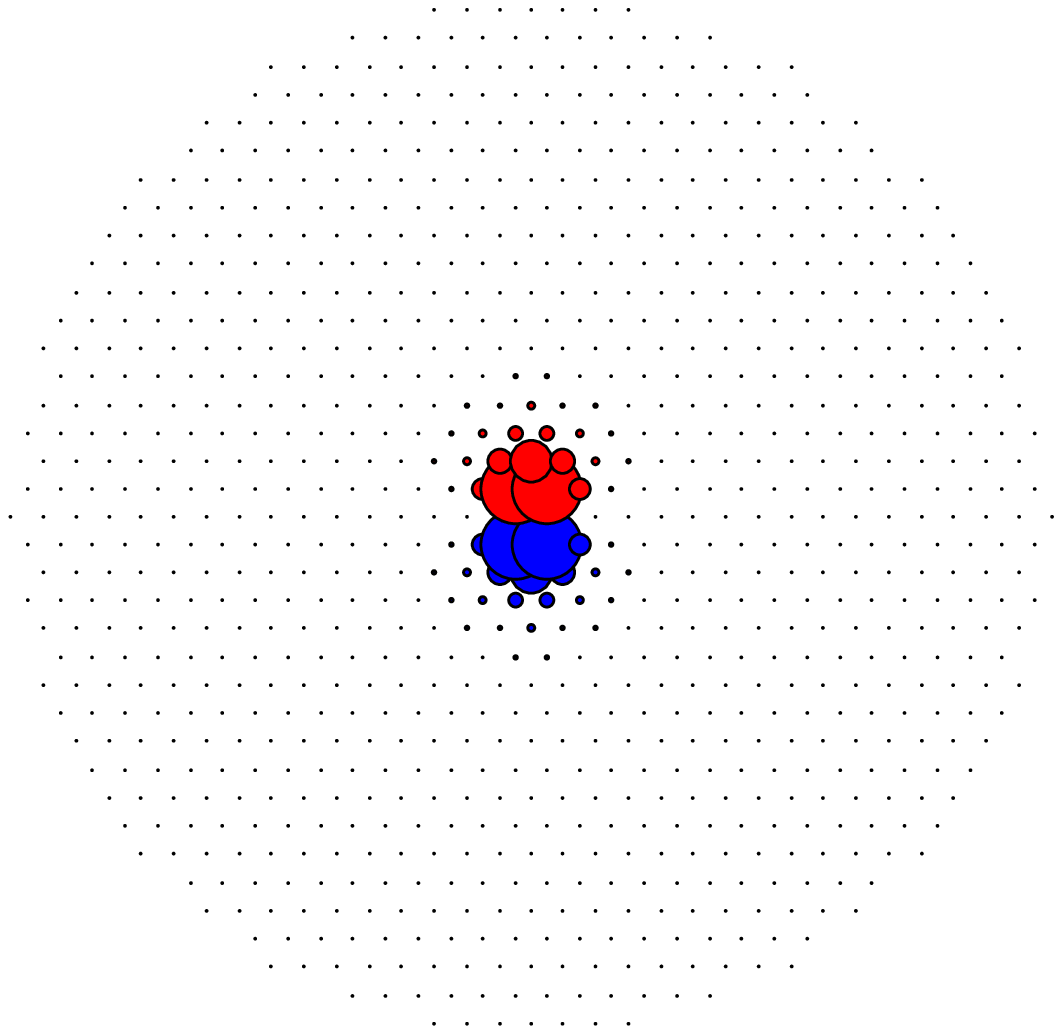}
\end{minipage}
\begin{minipage}{0.36\hsize}
\epsfxsize0.7\hsize\epsffile{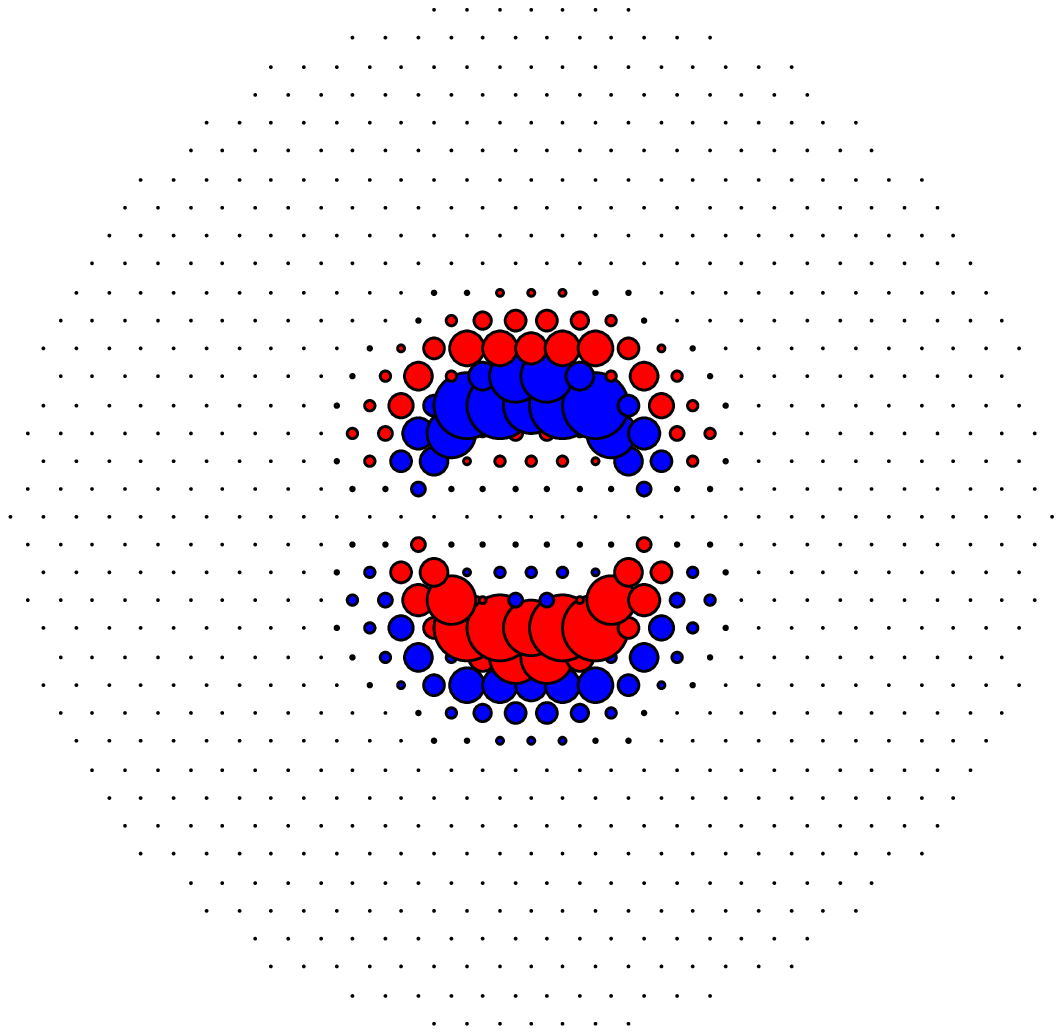}
\end{minipage}
}
%\end{tabular}
\caption{ 
\label{Fig3}
Out of phase magnetic field response 
for the case of an $m=-1, p=1$ vortex (left) and for
the case of an $m=-2, p=0$ vortex (right) 
to an {\em a.c.} electric field with $\omega = 0.2 \slDelta$
polarized in $x$-direction in the limit of high $\kappa $.
Red denotes negative values for the response, blue positive.
}
\end{figure}

In the high-$\kappa $ limit we can calculate the {\em a.c.} corrections
to the local magnetic field
in the vortex core directly from Eq.~\eqref{vecpot}, using
\begin{equation}
\delta \vec{B}(\vec{R},t) =\grad \times \delta \vec{A}(\vec{R},t) .
\end{equation}
We compare again the singly quantized and the doubly quantized vortex 
structures in Fig.~\ref{Fig3}. The time-dependent oscillating magnetic field
adds to the static, time-independent magnetic field that penetrates
the vortex in a quantized manner. We find that the dynamical magnetic
field response is predominantly out of phase, and resembles the patterns
expected for an oscillation of the magnetic field lines perpendicular to
the polarization of the external electric field. 
For the doubly quantized vortex, shown on the right hand side
in Fig.~\ref{Fig3}, the magnetic response is concentrated to
the region of overlap between the two time-reversed order 
parameter components.

\section{Conclusions}

We have discussed the electromagnetic response of a chiral, spin-triplet
superconducting vortex for two vortex structures, a singly quantized vortex
and a doubly quantized vortex. 
The vortex response is nonlocal and
largely determined by the response of the vortex-core states.
We have performed dynamically 
self consistent calculations, taking into account that
transitions involving the vortex-core states, and their coupling to the
collective motion of the condensate are closely related to the
dynamical response of the order parameter, self energies, induced fields,
excitation spectra and distribution functions. 

The vortex response roughly consists of an oscillation of the magnetic field lines
perpendicular to the electric field polarization, and an oscillation
of a charge dipole parallel to the electric field polarization. Both
the charge response and the magnetic response are linked to a current flow
pattern that show absorptive response patterns in the vortex core. 
Particularly interesting is the response for the doubly quantized vortex.
Here we find that the important region is the region where the time-reversed
order parameter phases overlap, which happens away from the vortex center
in a `domain wall' region. 

Based on our results, we see interesting future extensions of our work in 
the following directions: (i) more detailed study of the collective
order parameter response for the doubly quantized vortex; here
in particular the `optical' mode (oscillation of the two vortex
components against each other) is of interest; (ii) to include interactions
between the vortices by performing calculations on a vortex lattice; 
and (iii) to study the interplay between the charge response and the magnetic
response in a vortex lattice.

%\section*{Acknowledgements}
\ack
We acknowledge support from the National Science Foundation
DMR-0805277 (JAS) and from the Center for Functional Nanostructures of
the DFG (ME). ME also acknowledges
the hospitality of the Aspen Center for Physics.


\section*{References}
\begin{thebibliography}{10}

\bibitem{aaa57}
Abrikosov A A 1957 {\it Zh. Eksperim. Teor. Fiz.} {\bf 32} 1442
[1957 {\it Soviet Phys. JETP} {\bf 5} 1174]

\bibitem{lar79}
Larkin A I and Ovchinnikov Yu N
1978 {\it Pis'ma Zh. Eksp. Teor. Fiz.} {\bf 27} 301
[1978 {\it JETP Lett.} {\bf 27} 280]; 
1979 {\it J. Low. Temp. Phys.} {\bf 34} 409;
1980 {\it Zh. Eksp. Teor. Fiz.} {\bf 80} 2334
[1980 {\it Sov. Phys. JETP} {\bf 53} 1221];
Ovchinnikov Yu N
1980 {\it Zh. Eksp. Teor. Fiz.} {\bf 79} 1825
[1980 {\it Sov. Phys. JETP} {\bf 52} 923] 

\bibitem{degen64}
de Gennes P G and Matricon J 1964 {\it Rev. Mod. Phys.} {\bf 36} 45

\bibitem{gor73}
Gor'kov L P and Kopnin N B
1973 {\it Zh. Eksp. Teor. Fiz.} {\bf 64} 356
[1973 {\it Sov. Phys. JETP} {\bf 37} 183]; 
1973 {\it Zh. Eksp. Teor. Fiz.} {\bf 65} 396
[1974 {\it Sov. Phys. JETP} {\bf 38} 195] 

\bibitem{LO73}
Larkin A I and Ovchinnikov Yu N
1973 {\it Zh. Eksp. Teor. Fiz.} {\bf 64} 1096
[1973 {\it Sov. Phys. JETP} {\bf 37} 557]; 
1973 {\it Zh. Eksp. Teor. Fiz.} {\bf 65} 1704
[1074 {\it Sov. Phys. JETP} {\bf 38} 854];
1974 {\it Zh. Eksp. Teor. Fiz.} {\bf 66} 1100
[1974 {\it Sov. Phys. JETP} {\bf 39} 538]

\bibitem{GK75}
%vortex motion review
Gor'kov L P and Kopnin N B
1975 {\it Usp. Fiz. Nauk} {\bf 116} 413
[1976 {\it Sov. Phys.-Usp.} {\bf 18} 496].

\bibitem{bar65}
Bardeen J and Stephen M J 1965 {\it Phys. Rev.} {\bf 140} A1197

\bibitem{noz66}
%The motion of flux lines in type II superconductors
Nozi{\`e}res P and Vinen W F 1966 {\it  Philos. Mag.} {\bf 14} 667;
Vinen W F and Warren A C 1967 {\it Proc. Phys. Soc. London} {\bf 91} 409

\bibitem{sch66}
Schmid A 1966 {\it  Phys. Konden. Mater.} {\bf 5} 302;
Gor'kov L P and Kopnin N B
1971 {\it Zh. Eksp. Teor. Fiz.} {\bf 60} 2331
[1971 {\it Sov. Phys. JETP} {\bf 33} 1251]; 
Hu C-R and Thompson R S 1972 {\it Phys. Rev. B} {\bf 6} 110

\bibitem{LO75b}
Larkin A I and Ovchinnikov Yu N
1975 {\it Zh. Eksp. Teor. Fiz.} {\bf 68} 1915
[1976 {\it Sov. Phys. JETP} {\bf 41} 960]; 
1977 {\it Zh. Eksp. Teor. Fiz.} {\bf 73} 299
[1977 {\it Sov. Phys. JETP} {\bf 46} 155] 

\bibitem{LO76}
Larkin A I and Ovchinnikov Yu N 
1976 {\it Pis'ma Zh. Eksp. Teor. Fiz.} {\bf 23} 210
[1976 {\it JETP Lett.} {\bf 23} 187]


\bibitem{lar86}
%A.~I. Larkin and Y.~N. Ovchinnikov,  in {\em Modern Problems in Condensed
%Matter Physics}, edited by D. Langenberg and A. Larkin (Elsevier Science
%Publishers, Amsterdam, 1986).
Larkin A I and Ovchinnikov Yu N 1986
in {\it Nonequilibrium Superconductivity}, edited by D.N. Langenberg and A.I. Larkin (Elsevier, New York, 1986).

\bibitem{kop76}
Kopnin N B and Kravtsov  V E
1976 {\it Pis'ma Zh. Eksp. Teor. Fiz.} {\bf 23} 631
[1976 {\it JETP Lett.} {\bf 23} 578];
1976 {\it Zh. Eksp. Teor. Fiz.} {\bf 71} 1644
[1976 {\it Sov. Phys. JETP} {\bf 44} 861]

\bibitem{dor92}
%Vortex motion and the Hall effect in type-II superconductors: A time-dependent Ginzburg-Landau theory approach
Dorsey A T 1992 {\it Phys. Rev. B} {\bf 46} 8376

\bibitem{kop93}
Kopnin N B, Ivlev B I and Kalatsky V A 1992 {\it  Pis'ma Zh. Eksp. Teor. Fiz.} {\bf 55} 717 [1992 {\it JETP Lett.} {\bf 55} 750];
1993 {\it J. Low Temp. Phys.} {\bf 90} 1

\bibitem{kop94}
Kopnin N B 1994 {\it Pis'ma Zh. Eksp. Teor. Fiz.} {\bf 60} 123
[1994 {\it JETP Lett.} {\bf 60} 130];
1994 {\it J. Low Temp. Phys.} {\bf 97} 157

\bibitem{kop95}
%Flux-flow Hall effect in clean type-II superconductors
Kopnin N B and Lopatin A V 1995 {\it Phys. Rev. B} {\bf 51} 15291

\bibitem{LO95}
%Hall effect in type-II superconductors
Larkin A I and Ovchinnikov Yu N 1995 {\it Phys. Rev. B} {\bf 51} 5965

\bibitem{kop96}
%Hall effect in moderately clean superconductors and the transverse force on a moving vortex
Kopnin N B 1996 {\it Phys. Rev. B} {\bf 54} 9475

\bibitem{hou98}
%Quasiclassical approach to transport in the vortex state and the Hall effect
Houghton A and Vekhter I 1998 {\it Phys. Rev. B} {\bf 57} 10831

\bibitem{car64}
Caroli C, deGennes P G and Matricon J 1964 {\it Phys. Lett.} {\bf 9} 307

\bibitem{bar69}
Bardeen J, K\"ummel R, Jacobs A E and Tewordt L 
1969 {\it Phys. Rev.} {\bf 187} 556

\bibitem{rai96}
Rainer D, Sauls J A and Waxman D 1996 {\it  Phys. Rev. B} {\bf 54} 10094

\bibitem{Blatter94}
Blatter G, Feigel'man M V, Geshkenbein V B, Larkin A I and 
Vinokur V M 
1994 {\it Rev. Mod. Phys.} {\bf 66} 1125

\bibitem{Bill_vortex}
Chen B, Halperin W P, Guptasarma P, Hinks D G, Mitrovi{\'c} V F, Reyes A P and Kuhns P L 
2007 {\it Nature Physics} {\bf 3} 239

\bibitem{karrai92}
Karra{\"i} K, Chai E J, Dunmore F, Liu S H, Drew H D, Li Q, Fenner D B,
Zhu Y D and Zhang F-Ch
%{\it et al.}, 
1992 {\it Phys. Rev. Lett.} {\bf 69} 152

\bibitem{eldridge95}
Eldridge J E, Dressel M, Matz D J, Gross B, Ma Q Y and Hardy W N 
%{\it et al.}, 
1995 {\it Phys. Rev. B} {\bf 52} 4462

\bibitem{mae07}
Maeda A, Kitano H, Kinoshita K, Nishizaki T, Shibata K and Kobayashi N
%{\it et al.,} 
2007 {\it J. Phys. Soc. Jap.} {\bf 76} 094708

\bibitem{jan92}
Jank{\'o} B and Shore J 1992 {\it Phys. Rev. B} {\bf 46} 9270

\bibitem{zhu93}
Zhu Y D, Zhang J C and Drew H D 1993 {\it Phys. Rev. B} {\bf 47} 586

\bibitem{hsu95}
Hsu T 1993 {\it Physica C} {\bf 213} 305;
1995 {\it Phys. Rev. B} {\bf 52} 9178

\bibitem{lar98}
%Resistance of layered superclean superconductors at low temperatures 
Larkin A I and Ovchinnikov Yu N 1998 {\it Phys. Rev. B} {\bf 57} 5457

\bibitem{kou99}
Koulakov A A and Larkin A I 1999 {\it Phys. Rev. B} {\bf 59} 12021;
1999 {\it Phys. Rev. B} {\bf 60} 14597

\bibitem{atk99}
Atkinson W A and MacDonald A H 1999 {\it Phys. Rev. B} {\bf 60} 9295

\bibitem{kop97}
Kopnin N B and Volovik G E 1997 {\it Phys. Rev. Lett.} {\bf 79} 1377

\bibitem{eil68}
Eilenberger G 1968 {\it Zeit.f. Physik} {\bf 214} 195

\bibitem{lar68}
Larkin A I and Ovchinnikov Yu N 1968 {\it Zh. Eksp. Teor. Fiz.} {\bf 55} 2262
[1969 {\it Sov. Phys. JETP} {\bf 28} 1200]

\bibitem{eli71}
Eliashberg G M 1971 {\it Zh. Eksp. Teor. Fiz.} {\bf 61} 1254
[1971 {\it Sov. Phys. JETP} {\bf 34} 668]

\bibitem{schmid75}
Schmid A and Sch\"on G 1975 {\it Journ. Low. Temp. Phys.} {\bf 20} 207

\bibitem{esc97}
Eschrig M 1997 PhD Thesis, Univ. Bayreuth;
Eschrig M and Rainer D 1996 {\it Proc. LT21, Czech. J. Phys.} {\bf 46} 987

\bibitem{esc99}
Eschrig M, Sauls J A and Rainer D 1999 {\it Phys. Rev. B} {\bf 60} 10447

\bibitem{esc01}
Eschrig M, Sauls J A , Burkhardt H and Rainer D 2001
{\it Fermi Liquid Superconductivity: Concepts, Equations, Applications},
in { ``High-T$_c$ Superconductors and Related Materials -
Materials Science, Fundamental Properties, and Some Future Electronic
Applications''}, Proc. of the NATO Adv. Study Inst.,
NATO Science Partnership-Sub-Series 3, vol. 86,
edited by S.-L. Drechsler and T. Mishonov,
Kluwer Academic and Springer Netherland,
ISBN 0-7923-6872-X, pp. 413-446

\bibitem{esc02}
Eschrig M, Rainer D and Sauls J A 2002
{\it Vortex Core Structure and Dynamics in Layered Superconductors},
in ``Vortices in Unconventional Superconductors and Superfluids'',
edited by R.P. Huebener, N. Schopohl, and G.E. Volovik,
Springer Verlag, ISBN 3-540-42336-2, pp. 175-202


\bibitem{kho95}
Khomskii D I and Freimuth A 1995 {\it Phys. Rev. Lett.} {\bf 75} 1384

\bibitem{fei95}
Feigel'man M V, Geshkenbein V B, Larkin A I and Vinokur V M 1995
{\it JETP Lett.} {\bf 62}  834

\bibitem{bla96}
Blatter G, Feigel'man M, Geshkenbein V, Larkin A I and van Otterlo A 1996
%{\it et~al.}, 
{\it Phys. Rev. Lett.} {\bf 77}  566

\bibitem{kol01}
Kol\'a\v{c}ek J, Lipavsk\'y P and Brandt E H 2001 {\it Phys. Rev. Lett.} {\bf 86} 312

\bibitem{kum01}
Kumagai K-i, Nozaki K and Matsuda Y 2001 {\it Phys. Rev. B} {\bf 63} 144502

\bibitem{che02}
Chen Y, Wang Z D, Zhu J-X and Ting C S 2002 {\it Phys. Rev. Lett.} {\bf 89} 217001

\bibitem{shi02}
\v{S}im\'anek E 2002 {\it Phys. Rev. B} {\bf 65} 184524

\bibitem{lip02}
Lipavsk\'y P, Kol\'a\v{c}ek J, Morawetz K
and Brandt E H 2002 {\it Phys. Rev. B} {\bf 65} 144511

\bibitem{mac03}
Machida M and Koyama T 2003 {\it Phys. Rev. Lett.} {\bf 90} 077003

\bibitem{kna05}
Knapp D, Kallin C, Ghosai A and  Mansour S 2005 {\it Phys. Rev. B} {\bf 71} 064504

\bibitem{zhu06}
Zhu B-H, Zhou S-P, Shi Y-M, Zha G-Q, and Yang K 2006 {\it Phys. Rev. B} {\bf 74} 014501

\bibitem{zha08}
Zhao H-W, Zha G-Q, Zhou S-P and Peeters F M 2008 {\it Phys. Rev. B} {\bf 78} 064505

\bibitem{Sorokin49}
Sorokin V S 1949 {\it  Sov. Phys. JETP } {\bf 19} 553

\bibitem{Gorter34}
Gorter C J and Casimir H B G  1934 {\it Phys. Z.} {\bf 35} 963; 
1934 {\it Z. Tech. Phys.  (Leipzig)} {\bf 15} 539; 
1934 {\it Physica (The Hague)} {\bf 1} 306

\bibitem{Bardeen54}
Bardeen J 1954 {\it Phys. Rev.} {\bf 94} 554

\bibitem{Ginzburg50}
Ginzburg V L and Landau L D 1950 {\it Zh. Eksp. Teor. Fiz.} {\bf 20}, 1064

\bibitem{Rickayzen69}
Rickayzen G 1969 {\it J. Phys. C} {\bf 2} 1334

\bibitem{Adkins68}
Adkins C J and Waldram J R 1968 {\it Phys. Rev. Lett.} {\bf 21} 76

\bibitem{Bopp37}
Bopp F 1937 {\it Z. Phys.} {\bf 107} 623

\bibitem{London50}
London F 1950 {\it Superfluids} (Willey, New York, 1950), Vol. I, Sec. 8.

\bibitem{Vijfeijken64}
van Vijfeijken A G and Staas F S 1964 {\it Phys. Lett.} {\bf 12} 175

\bibitem{Thomas}
Thomas L H 1927 {\it Proc. Cambridge Philos. Soc.} {\bf 23} 542;
Fermi E 1928 {\it Z. Phys.} {\bf 48} 73;
%\bibitem{Jakeman67}
Jakeman E and Pike E R 1967 {\it Proc. Phys. Soc.} {\bf 91} 422

\bibitem{art79}
Artemenko S and Volkov A 1979
{\it Usp. Fiz. Nauk} {\bf 128} 3
[1979 {\it Sov. Phys.-Usp.} {\bf 22}  295]

\bibitem{Sauls09}
Sauls J A and Eschrig M 2009 Vortices in chiral, spin-triplet superconductors
and superfluids. {\it New Journal of Physics}
Focus Issue on {\it Superconductors with Exotic Symmetries},
M. Rice and M. Sigrist, editors. arXiv:0903.0011

\bibitem{ser83}
Serene J W and Rainer D 1983 {\it Phys. Rep.} {\bf 101} 221

\bibitem{rai95}
Rainer D and Sauls J A 1995 in {\em Superconductivity: From Basic Physics to
New Developments}, edited by P.~N. Butcher and Y. Lu (World Scientific,
Singapore), pp.\ 45--78.

\bibitem{rai94}
Eschrig M, Heym J and Rainer D 1994
{\it J. Low Temp. Phys.} {\bf 95} 323

\bibitem{esc99fluk}
Eschrig M, Rainer D and Sauls J A 1999 {\it Phys. Rev. B} {\bf 59} 12095
Appendix C

\bibitem{Bill}
Halperin W P, Mukhopadhyay S, Mounce A M, Oh S, Reyes A P, Kuhns P, 
Takagi H and Uchida S 2009
{\it Spatial distribution of internal magnetic field in High-$T_c$ superconductors with pancake vortices}, Abstr. X34.00002, 2009 APS meeting

\bibitem{kel64}
Keldysh L V 1964 {\it Zh. Eksp. Teor. Fiz.} {\bf 47} 1515
[1965 {\it Sov. Phys. JETP } {\bf 20} 1018]

\bibitem{kop78a}
Kopnin N B 1978
{\it Pis'ma Zh. Eksp. Teor. Fiz.} {\bf 27} 417
[1978 {\it JETP Lett.} {\bf 27} 390]

\bibitem{esc00}
Eschrig M 2000 {\it Phys. Rev. B} {\bf 61} 9061

\end{thebibliography}
\end{document}